\documentclass[lettersize,journal]{IEEEtran}
\usepackage{algorithm}
\usepackage{stfloats}
\usepackage{url}
\usepackage{verbatim}
\usepackage{cite}
\usepackage{amsmath,amssymb,amsfonts}
\usepackage{algorithmic}
\usepackage{gensymb}
\usepackage{amsthm}

\usepackage{booktabs}
\usepackage{array}
\usepackage{adjustbox}
\usepackage{tabularray}
\usepackage{diagbox}

\usepackage{graphicx,subfigure}
\usepackage{textcomp}
\usepackage{xcolor}
\usepackage{flushend}

\begin{document}

\title{GDSG: Graph Diffusion-based Solution Generator for Optimization Problems in MEC Networks}

\author{Ruihuai~Liang,~\IEEEmembership{Student Member,~IEEE,} Bo~Yang,~\IEEEmembership{Member,~IEEE,} Pengyu~Chen, Xuelin Cao,~\IEEEmembership{Member,~IEEE,} Zhiwen Yu,~\IEEEmembership{Senior Member,~IEEE,}   M\'erouane Debbah,~\IEEEmembership{Fellow,~IEEE}, Dusit Niyato,~\IEEEmembership{Fellow,~IEEE}, \\ H. Vincent Poor,~\IEEEmembership{Life Fellow,~IEEE},  and Chau~Yuen,~\IEEEmembership{Fellow,~IEEE}  % <-this % stops a space
\thanks{R. Liang, B. Yang, and P. Chen are with the School of Computer Science, Northwestern Polytechnical University, Xi'an, Shaanxi, 710129, China. 

Z. Yu is with the School of Computer Science, Northwestern Polytechnical University, Xi'an, Shaanxi, 710129, China, and Harbin Engineering University, Harbin, Heilongjiang, 150001, China.

X. Cao is with the School of Cyber Engineering, Xidian University, Xi'an, Shaanxi, 710071, China. 

M. Debbah is with the Center for 6G Technology, Khalifa University of Science and Technology, P O Box 127788, Abu Dhabi, United Arab Emirates. 

D. Niyato is with the College of Computing and Data Science, Nanyang Technological University, Singapore.

H. V. Poor is with the Department of Electrical and Computer Engineering, Princeton University, Princeton, NJ 08544, USA.

C. Yuen is with the School of Electrical and Electronics Engineering, Nanyang Technological University, Singapore. 

 (\textit{Corresponding author:} Bo Yang)
 }
%\thanks{Manuscript received April 19, 2005; revised August 26, 2015.}
}

% The paper headers
\markboth{Journal of \LaTeX\ Class Files,~Vol.~XX, No.~XX, December~2024}%
{Shell \MakeLowercase{\textit{et al.}}: A Sample Article Using IEEEtran.cls for IEEE Journals}

% \IEEEpubid{0000--0000/00\$00.00~\copyright~2021 IEEE}
% Remember, if you use this you must call \IEEEpubidadjcol in the second
% column for its text to clear the IEEEpubid mark.

\maketitle

\begin{abstract}

Optimization is crucial for the efficiency and reliability of multi-access edge computing (MEC) networks. Many optimization problems in this field are NP-hard and do not have effective approximation algorithms. Consequently, there is often a lack of optimal (ground-truth) data, which limits the effectiveness of traditional deep learning approaches. Most existing learning-based methods require a large amount of optimal data and do not leverage the potential advantages of using suboptimal data, which can be obtained more efficiently. \textcolor{black}{To illustrate this point, we focus on the multi-server multi-user computation offloading (MSCO) problem, a common issue in MEC networks that lacks efficient optimal solution methods. In this paper, we introduce the graph diffusion-based solution generator (GDSG), designed to work with suboptimal datasets while still achieving convergence to the optimal solution with high probability.}
We reformulate the network optimization challenge as a distribution-learning problem and provide a clear explanation of how to learn from suboptimal training datasets. We develop GDSG, a multi-task diffusion generative model that employs a graph neural network (GNN) to capture the distribution of high-quality solutions. Our approach includes a straightforward and efficient heuristic method to generate a sufficient amount of training data composed entirely of suboptimal solutions.
In our implementation, we enhance the GNN architecture to achieve better generalization. Moreover, the proposed GDSG can achieve nearly 100\% task orthogonality, which helps prevent negative interference between the discrete and continuous solution generation training objectives. We demonstrate that this orthogonality arises from the diffusion-related training loss in GDSG, rather than from the GNN architecture itself. Finally, our experiments show that the proposed GDSG outperforms other benchmark methods on both optimal and suboptimal training datasets. Regarding the minimization of computation offloading costs, GDSG achieves savings of up to 56.62\% on the ground-truth training set and 41.06\% on the suboptimal training set compared to existing discriminative methods\footnote{The MSCO datasets are open-sourced at https://dx.doi.org/10.21227/386p-7h83, and the GDSG algorithm codes at https://github.com/qiyu3816/GDSG.}.
\end{abstract}


\section{Introduction}
\IEEEPARstart{I}{n} various cutting-edge scenarios of multi-access edge computing (MEC) \cite{mec2017mao,mec2017mach}, optimization demands are ubiquitous, such as \textcolor{black}{trajectory optimization in uncrewed aerial vehicle (UAV) networks \cite{coletta20232,zhan2022energy,multiUAV2021chai,multiUAV2021zhao,yb_uav,uav2024duong,uav2025Duong}}, utility and latency optimization in edge networks for federated learning (FL) \cite{edgeFL2022nguyen,edgeFL2022karakoc,yb_fl}, sum-rate and coverage optimization in reconfigurable intelligent surface (RIS)-enhanced networks \cite{cao-jsac,ris2020multiuser,feng2023resourceAll,mei2023joint,ris2024xueyao}, efficiency optimization in Internet-of-Vehicles (IoV) \cite{offloading2024MEC_IoV,dai2021asynchronous,yb_edge_intelligence}, and other resource allocation tasks \cite{he2023deepscheduler,rusek2020routenet,yb_iiot,jiang2019deep,liang2023multi,zhang2023gnn,cxl_leo}. Despite the significant advancements in numerical algorithms and artificial intelligence (AI) methods, a key challenge persists: a large number of these problems are NP-hard, and there are currently no efficient approximation algorithms available. This limitation not only restricts algorithms to producing suboptimal results, despite extensive manual design efforts, but also hinders the potential of AI-based methods, which typically rely on large datasets consisting of optimally solved examples that are inherently difficult to obtain in practice. 

Fortunately, the success of generative AI has demonstrated its ability to generalize beyond the training data, offering a more robust AI approach to data quality. Large language models (LLMs) trained on a limited amount of text can generate novel responses to diverse queries \cite{star2024zelikman,pmlr-v235-ma24m} rather than simply reassembling the training data. Similarly, generative diffusion models (GDMs) trained on image datasets can create significantly different images under various conditions \cite{dream2024du,controlNet2023}, rather than simply recalling training samples. \textcolor{black}{\textit{Based on this evidence, we propose the hypothesis that generative AI models can solve complex network optimization problems, even in the absence of an exclusively optimal dataset.}}

\subsection{Motivation and Contribution}
Many optimization problems in MEC networks lack efficient solution methods. \textcolor{black}{Existing commercial and open-source solvers, such as Gurobi \cite{gurobi}, Mosek \cite{mosek}, CPLEX \cite{cplex}, and GEKKO \cite{beal2018gekko}, often face excessively high algorithmic complexity during iterative optimization processes. For common challenges, such as non-convex optimization and mixed-integer programming, these solvers can typically guarantee efficiency only in low-dimensional cases, sometimes with fewer than $10$ dimensions. Furthermore, many network optimization problems still lack mature and effective algorithms, leaving them without ready-made solvers. The limitations of traditional solvers in handling high-complexity problems highlight the necessity for AI-based methods. However, generating a large number of optimal solution samples for training purposes is prohibitively expensive, both in terms of time and computational resources.} In such scenarios, deriving suboptimal solutions often proves easier than obtaining a sufficient number of optimal solutions, which can be time-consuming and labor-intensive. For example, the multi-objective nature of certain optimization problems can be effectively addressed using divide-and-conquer strategies. This approach enables the development of suboptimal solutions that are straightforward to implement and computationally efficient, making them suitable for constructing large-scale datasets. Although these suboptimal instances may not offer significant advantages for current AI techniques, they provide an opportunity to explore the potential of GDM in generating solution distributions and achieving improved outcomes through parallel sampling.

To validate our earlier hypothesis, we employ generative AI to extract insights from suboptimal datasets to develop solutions for MEC optimization problems. In this context, we consider the feasible solution space of a given input as a distribution of potential solutions, with the optimal solution being the one that has the highest probability within this distribution. We then parameterize the resulting solution distribution to facilitate effective distribution learning. Following this step, we can simultaneously sample from the parameterized distribution generated from the input, along with an approximation of the optimal solution. \textcolor{black}{Generative models possess the ability to learn a target distribution and sample from it. Among these models, GDM stands out for the balance of performance and efficiency \cite{cao2024survey}. Consequently, we utilize GDM to learn the parameterized solution distribution and to further sample high-quality solutions from it.}

\textcolor{black}{Given that many problems in MEC can be formulated as graph optimization problems, such as topology design and routing scheduling \cite{rusek2020routenet, he2023deepscheduler, zhang2023gnn, chen2021gnn, mei2023joint}. To address this, we focus on the multi-server multi-user computation offloading (MSCO) problem and utilize graph diffusion techniques to find a solution. We propose a solution generator called the graph diffusion-based solution generator (GDSG). To train GDSG, we first create a suboptimal training set using heuristic methods and then employ a conditional generation mechanism. During the solving phase, we use the model to perform parallel sampling of high-quality solutions, which helps us approximate the optimal solution.}

To the best of our knowledge, \textit{this is a pioneering attempt to solve complex optimization problems that take into account the training data limitations in MEC networks}. The key contributions of our work are outlined below.
\begin{itemize}
    \item \textcolor{black}{We explore the potential of utilizing suboptimal data samples and clarify the reasoning for learning the solution distribution instead of relying solely on GDM with a large number of optimal solution samples. Our theoretical framework discusses the convergence toward the optimal solution and summarizes the conditions required for convergence.} 
    \item \textcolor{black}{We present GDSG, a method trained on suboptimal data for graph optimization, demonstrating the capability to generate high-quality solutions with specific inputs through parallel sampling. This approach offers a more efficient network optimization solution than numerical algorithms while also incurring lower data costs compared to traditional AI methods, all while ensuring optimal convergence.}
    \item We provide some practical insights from engineering implementation aspects. By explicitly enhancing the graph neural network (GNN) gating mechanism, we improve the model’s generalization ability across different graph scales. Additionally, our analysis of multi-task training reveals the orthogonality introduced by the diffusion model’s loss objective, leading to negligible interference between tasks. 
\end{itemize}

\subsection{Related works}
\subsubsection{Learning-based optimization}
Current learning-based approaches to network optimization can be classified based on their reliance on training data. The first category utilizes ground-truth datasets \cite{yang2020computation,liang2023multi,jiang2019deep,zhang2023gnn,chen2021gnn,multiUAV2021chai,cao-jsac}, either created from scratch using standard solvers, brute force techniques, or derived from open-source datasets. However, open-source datasets often have limited coverage, and the use of solvers based on problem-specific assumptions, along with the high cost of dataset construction, makes this approach impractical for new scenarios. The second category seeks to minimize reliance on already solved datasets \cite{dai2021asynchronous,he2023deepscheduler,rusek2020routenet,edgeFL2022nguyen,multiUAV2021zhao,offloading2024MEC_IoV}. This approach primarily uses reinforcement learning, where the model is guided to discover improved policies through the careful design of rewards and losses. Nevertheless, this method heavily depends on expert knowledge and requires adjustments when the scenarios change.

\subsubsection{Graph diffusion and optimization}
Diffusion models are a common type of generative model that involve introducing noise into data and then learning to remove that noise. This approach enables the model to predict and progressively restore the desired data distribution by repeatedly correcting for the noise. One notable example is the denoising diffusion probabilistic model (DDPM)~\cite{ho2020denoising}, which can produce distributions based on specific conditions~\cite{ho2022classifier,dhariwal2021diffusion}. 

Graph diffusion, on the other hand, involves learning and generating non-Euclidean graph-structured data using GNNs \cite{austin2021structured}. This approach has demonstrated success in addressing combinatorial optimization problems such as the traveling salesman problem (TSP) and the maximum independent set problem (MIS), yielding impressive results in both \cite{sun2023difusco} and \cite{li2024distribution}. Specifically, the T2T method \cite{li2024distribution} improves the sampled solutions by incorporating gradient-guided generation based on DIFUSCO \cite{sun2023difusco}. Thus far, graph diffusion models have proven advantageous by explicitly modeling graph structures to learn features of multi-modal distributions.

Recent studies \cite{du2024enhancing,liang2024diffsggenerativesolvernetwork,wang2024generativeaibasedsecure} have explored the integration of diffusion models with network optimization. However, this research mainly focuses on practical applications in various optimization problems without considering data robustness. Additionally, with new combinations of GDM and reinforcement learning in the AI-community, some studies \cite{sun2024generativeaideepreinforcement,du2024D2SAC,liu2024deep} utilize GDM as a more effective policy network in deep reinforcement learning (DRL) to tackle network optimization challenges. Unlike the related works mentioned above, the proposed GDSG approach provides an effective implementation of graph-level modeling for solving complex problems via multi-task optimization in MEC networks. The suboptimal datasets can be efficiently created using basic heuristic algorithms, which ensure the quality of the training data for GDSG. \textcolor{black}{Our research specifically explores the potential of utilizing suboptimal datasets with GDMs, which can lead to robust performance outcomes. Our theoretical findings and engineering results provide GDSG with interpretable superior performance and reduce implementation barriers. This approach helps to eliminate confusion arising from unknown underlying principles and addresses engineering challenges associated with data scarcity.}

\section{System Model and Problem Formulation}\label{sec_problem}
\textcolor{black}{The advancement of various MEC systems \cite{edgeFL2022karakoc,edgeFL2022nguyen,admm2024cy,admm2022tnse,admm2024twc} has increased the demand for complex computational tasks, resulting in resource shortages on user devices. In this context, computation offloading has emerged as a crucial solution to address this issue.}

\subsection{System Model}
Our considered MSCO system consists of multiple edge servers and mobile users, where mobile users represent a unified abstraction of devices such as smartphones, drones, and vehicles. The number of servers and users in the system is variable, allowing for dynamic node mobility and the flexible joining or leaving of nodes based on task requirements. 

\subsubsection{System overview}
\begin{figure}[t]
\centerline{\includegraphics[width=3.35in, height=1.6in]{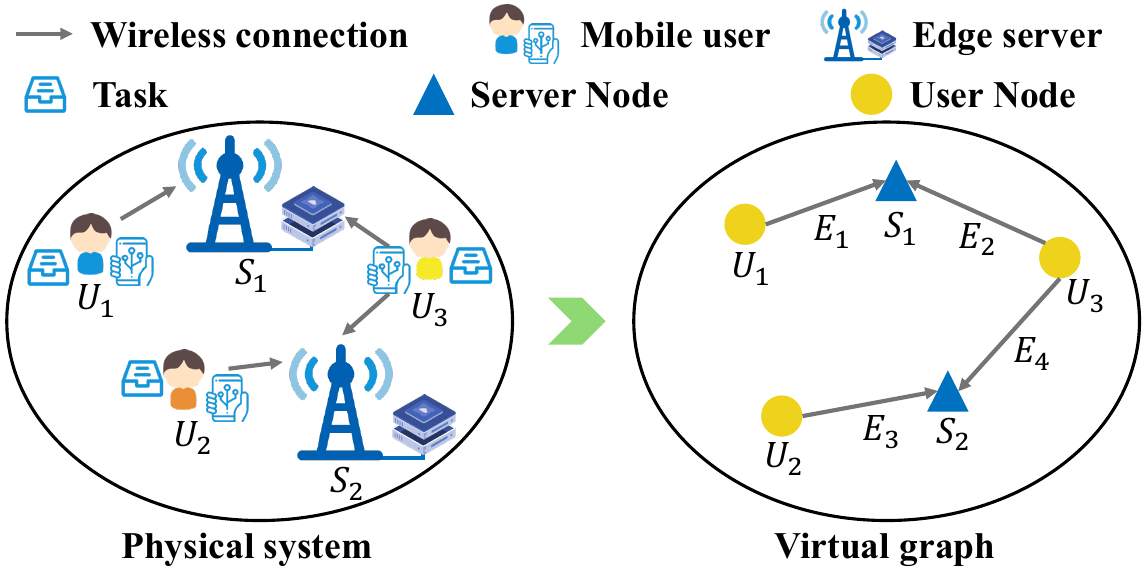}}
\setlength{\abovecaptionskip}{-0.1cm}
\caption{System model of multi-server and multi-user computation offloading.}
\label{fig_sys_model}
\vspace{-0.50cm}
\end{figure}

We formulate a binary computation offloading problem involving multiple servers and multiple users to minimize the weighted cost of delay and energy consumption, with the optimization variables of offloading decision and server computational resource allocation. As illustrated in Fig. \ref{fig_sys_model}, there exists a set of $K$ edge servers, denoted as $\mathcal{S}\!=\!\{S_1,...,S_K\}$, and a set of $M$ mobile users, denoted as $\mathcal{U}\!=\!\{U_1,...,U_M\}$ within the specified area. Due to the limited radio coverage of the edge servers, each user can only connect with a subset of edge servers. We assume that the uplink multiple access mechanism is orthogonal multiple access (OMA) \cite{yaacoub2011survey,salem2010overview} to ensure interference-free wireless transmission. 

Abstracting the problem to a graph structure, we obtain the virtual graph of Fig. \ref{fig_sys_model}. Let $\mathcal{G}(\mathcal{V},\mathcal{E})$ represent this directed acyclic graph, where $\mathcal{V}\!=\!\{\mathcal{U},\mathcal{S}\}$. All feasible wireless connections are considered as directed edges from users to servers $\mathcal{E}=\{E_i(U_j, S_k)|{\rm if}\ U_j\ {\rm has\ connection\ with}\ S_k\}$, where $L=|\mathcal{E}|$ and $i\in\{1,...,L\}, j\in\{1,...,M\}, k\in\{1,...,K\}$. 

\subsubsection{Optimization variables}
We can observe that the dimension of the optimization variable is the same as the number of edges. 
Let $\mathcal{D}=\{D_1,...,D_i,...,D_L\}$ be the binary offloading decision of the mobile users $\mathcal{U}$, where $D_i\in\{0,1\}$. So user $U_j$ offloads to $S_k$ if $E_i(U_j,S_k)$ exists and $D_i=1$, otherwise $U_j$ does not offload to $S_k$. If all offloading decisions for the edges associated with $U_j$ are 0, it means that the computation task of $U_j$ is executed locally rather than being offloaded. 

Let $\mathcal{A}=\{A_1,...,A_i,...,A_L\}$ be the computational resource allocation for the mobile users $\mathcal{U}$, where $A_i\in[0,1]$ represents the proportion of the total available computational resource from the server allocated to the user of $E_i$. Intuitively, we suppose that $\mathcal{D}$ and $\mathcal{A}$ determine the retention and weight of edges in the result, respectively.  

\subsubsection{Binary offloading}
During the offloading optimization, each user is assumed to have an indivisible computing task, either executed locally or offloaded to a MEC server for execution. To intuitively formalize the relationship between edges and nodes, we define a binary function, i.e.,  
\begin{equation}\label{eq_source_node_j}
\begin{aligned}
    \Upsilon(E_i,U_j)=\left\{
    \begin{matrix}
        1 & {\rm if }\ U_j \ {\rm is\ the\ source\ node\ of }\ E_i \\ 
        0 & {\rm otherwise}
    \end{matrix}
    \right. \quad \\
        \forall i\in\{1,...,L\},\ \forall j\in\{1,...,M\}&,
\end{aligned}
\end{equation}
to determine whether $U_j$ is the source node of the directed edge $E_i$ or not. Therefore, the outgoing edges of a user node $U_j$ should satisfy $\sum_{\Upsilon(E_i,U_j)=1} D_i\leq 1$. 

In addition, the total amount of computational resources allocated to the tasks offloaded to a server cannot exceed the available computational resources of the server. Therefore, we define a function 
\begin{equation}\label{eq_des_node_j}
\begin{aligned}
    \Omega(E_i,S_k)=\left\{
    \begin{matrix}
        1 & {\rm if }\ S_k \ {\rm is\ the\ destination\ node\ of }\ E_i \\ 
        0 & {\rm otherwise}
    \end{matrix}
    \right. \quad \\
        \forall i\in\{1,...,L\},\ \forall k\in\{1,...,K\}&,
\end{aligned}
\end{equation}
to determine whether $S_k$ is the destination node of the directed edge $E_i$ or not. As a result, the incoming edges of a server node $S_k$ should satisfy $\sum_{\Omega(E_i,S_k)=1} D_i A_i\leq 1$. 

\subsection{Weighted Cost of Delay and Energy Consumption}
%\subsubsection{Parameters of a user node for local computation}
Let $\mathbf{I}=\{I_1,...,I_L\}$, $\mathbf{V}=\{V_1,...,V_L\}$ and $\mathbf{F}=\{F^l_1,...,F^l_L\}$ be the input data sizes [bits], required computational resources [cycles] and available local computational resources [cycles per second] of the user tasks on each edge. Let $\mathcal{\alpha}=\{\alpha_1,...,\alpha_L\}$ represent the weight of the task cost, respectively, where $\alpha_i\in[0,1]$. So we have $\alpha\to 1$ if the delay cost is emphasized while the energy consumption cost is emphasized if $\alpha\to 0$. Clearly, the parameters related to local computation are the same for all outgoing edges of the same user node. Thus, we have 
\begin{equation}\label{eq_edges_same_source}
\begin{aligned}
I_i=&I_j, V_i=V_j, F^l_i=F^l_j, \alpha_i=\alpha_j,\ {\rm if\ E_i\ and\ E_j}\\ 
&{\rm have\ the\ same\ source\ node},\ \forall i,j\in\{1,...,L\}.
\end{aligned}
\end{equation}

\subsubsection{Local execution cost}
For $E_i,i\in\{1,...,L\}$, its local execution time is calculated as $\tau_i=\frac{V_i}{F^l_i}$. Then, its local execution energy consumption can be obtained according to the energy consumption model \cite{chen2014decentralized} as $\omega^l_i=\kappa (V_i)^2 I_i$, where $\kappa$ represents the energy efficiency parameter mainly depending on the chip architecture \cite{burd1996processor}. The weighted cost of local execution can be calculated as 
\begin{equation}\label{eq_local_cost}
C^l_i=\alpha_i \tau^l_i + (1-\alpha_i)\omega^l_i,\ \forall i\in\{1,...,L\}.
\end{equation}

\subsubsection{Offloading execution cost}
Let $\mathcal{H}=\{h_1,...,h_i,...,h_L\}, \ h_i\in[0,1]$ be the channel gain and $B$ be the channel bandwidth of the wireless connection represented by each edge. Let the power [W] of the uplink be $P_t$ with additive Gaussian white noise of zero mean and variance $N_0$. Then the uplink received signal-to-interference-plus-noise ratio (SINR) is given as $SINR_i=\frac{P_t (h_i)^2}{N_0+P_t\sum (h_j)^2}$, where $E_j$ and $E_i$ has the same server node. In this way, the transmission rate of $E_i$ is calculated as $r^u_i=B\log_2(1+SINR_i)$. The weighted cost of uploading transmission can be calculated as 
\begin{equation}\label{eq_trans_cost}
{C}^t_i=\alpha_i \frac{I_i}{r^u_i} + (1-\alpha_i)P_t\frac{I_i}{r^u_i},\ \forall i\in\{1,...,L\}.
\end{equation}

For the task execution on the server, let $F$ be the total available computational resource [cycles/second] of the server and $P_e$ be the processing power [W]. The weighted cost of fully-occupied offloading execution can be calculated as 
\begin{equation}\label{eq_offload_cost}
C^o_i=\alpha_i \frac{V_i}{F} + (1-\alpha_i)P_e\frac{V_i}{F},\ \forall i\in\{1,...,L\}.
\end{equation}
Therefore, the total offloading cost of $E_i$ can be calculated as $D_i(C^t_i+\frac{C^o_i}{A_i})$. 

%\subsubsection{Additional parameters}
Since the computing task has a maximum tolerable delay $\tau_i$ [seconds], let $\rho_i\in[0,1]$ represent the minimum computational resource allocation ratio required for offloading execution without timeout, and let $\psi_i\in\{0,1\}$ represent whether the local execution will timeout or not. Only if the offloading was able to process without a timeout, $\rho_i>0$. As a result, $\rho_i=0$ if $\frac{I_i}{r^u_i} + \frac{V_i}{F}>\tau_i$ and $\rho_i=\frac{V_i}{(\tau_i-I_i/r^u_i)F}$ if $\frac{I_i}{r^u_i} + \frac{V_i}{F}\leq\tau_i$. Furthermore, $\psi_i=0$ if the local execution will timeout, $\psi_i=1$ otherwise. Therefore, we use $(C^l_i,C^t_i,C^o_i,\rho_i,\psi_i)$ as the 5-dimensional edge feature in $\mathcal{G}$ to represent computing tasks, communication parameters, and resource settings. 

\subsection{Weighted Cost Minimization Problem}
According to Eqs. (\ref{eq_local_cost}), (\ref{eq_trans_cost}) and (\ref{eq_offload_cost}), the weighted cost minimization problem can be formulated as 
\begin{subequations}\label{eq_obj_func}
\begin{align}
    &  \mathbb{P}: \underset{\{\mathcal{D},\mathcal{A}\}}{{\rm min}}\ \sum^L_{i=1}(1-D_i)C^l_i+D_i(C^t_i+\frac{C^o_i}{A_i}), \notag \\
    &s.t.\ \mathbf{C1}: \ D_i\in\{0,1\}, \ \forall i \in \{1,...,L\}, \\
    &\ \ \ \ \ \mathbf{C2}: \ A_i\in[0,1], \ \forall i \in \{1,...,L\}, \\
    &\ \ \ \ \ \mathbf{C3}: \ \sum_{\Upsilon(E_i,U_j)=1} D_i\leq 1, \forall j\in\{1,...,M\}, \\
    &\ \ \ \ \ \mathbf{C4}: \ \sum_{\Omega(E_i,S_k)=1} D_i A_i\leq 1, \forall k\in\{1,...,K\}.
\end{align}
\end{subequations}

Let $\{\mathcal{D}^*, \mathcal{A}^*\}$ be the optimal solution for the given input parameters, the problem $\mathbb{P}$ can be transformed from the original mix-integer non-linear programming (MINLP) to a multi-task graph optimization. It is important to note that although $\mathbb{P}$ is formally an MINLP problem and the existing solvers cannot directly handle such a challenging graph-structured problem. For example, Gurobi and GEKKO solver only accept vector-based problem and constraint formulations, and the time cost for solving MINLP problems is excessively high. In this context, with the node features that represent node types (e.g., $1$ represents a server node and $0$ represents a user node) and edge features that explicitly denote objective and constraint, the graph-based formulation can provide rich information and can be transferable to alternative problems. 

\section{Problem Analysis}\label{sec_method_def}
In this section, we present the principles and concepts of solution distribution learning, explicitly highlighting the robustness of this learning objective with respect to solution quality, thereby solving the optimization problem in MEC networks from the suboptimal dataset. 

\subsection{Why and How to Learn the Solution Distribution?}
For a network optimization problem, existing methods \cite{yang2020computation,liang2023multi,jiang2019deep,zhang2023gnn,chen2021gnn,multiUAV2021chai,cao-jsac} generally obtain truth input-solution paired samples such as $(\mathbf{x}, \mathbf{y}^*), \mathbf{x}\in\mathcal{X}, \mathbf{y}^*\in\mathcal{Y}_{\mathbf{x}}$, $\mathcal{X}$ and $\mathcal{Y}_{\mathbf{x}}$ represent the input parameter space and the feasible solution space for a given $\mathbf{x}$, respectively. Inspired by \cite{qiu2022dimes}, we extend the definition of parametrized solution space and parametrize the solution space with a continuous vector $\boldsymbol{\theta}\in\mathbb{R}^N$, where $N$ represents the dimension of the optimization variable. This probability of each feasible solution $\mathbf{y}$ can be estimated as 
\begin{equation}\label{eq_parametrization}
p_{\boldsymbol{\theta}}(\mathbf{y}|\mathbf{x})\propto {\rm exp}(\sum^N_{i=1} y_i \theta_i)\ \ s.t.\ \mathbf{y}\in\mathcal{Y}_{\mathbf{x}},\ \mathbf{x}\in\mathcal{X},
\end{equation}
where $p_{\boldsymbol{\theta}}$ is an energy function indicating the probability of each solution over the feasible solution space. We first consider the discrete solution space that $\mathbf{y}$ is an $N$-dimensional vector with $y_i\in\{0,1\}$, and the higher value of $\theta_i$ denotes a higher probability that $y_i$ is consistent with $\mathbf{y}^*$. There exists $\boldsymbol{\theta}^*$ that is the optimal parametrization of the solution space for a given $\mathbf{x}\in\mathcal{X}$ if $\boldsymbol{\theta}^*$ satisfies $p_{\boldsymbol{\theta}^*}(\mathbf{y}^*|\mathbf{x})=1$. 

Accordingly, \(\boldsymbol{\theta}\) represents a solution distribution. Taking the discrete binary classification problem as an example, for the optimal solution \(\mathbf{y}^*\), it is parameterized using a binomial distribution. Specifically, \(\theta^*_{i0} = 0\) and \(\theta^*_{i1} = 1\) if \(y^*_i = 1\), while \(\theta^*_{i0} = 1\) and \(\theta^*_{i1} = 0\) if \(y^*_i = 0\). Here, \(\theta^*_{i0}\) represents the probability that \(y_i = 0\), and \(\theta^*_{i1}\) represents the probability that \(y_i = 1\). Therefore, \((\mathbf{x}, \mathbf{y}^*)\) can be replaced by \((\mathbf{x}, \boldsymbol{\theta}^*)\) for the purpose of solution learning.

Although learning \(\boldsymbol{\theta}^*\) can guarantee accurate results, it is clear that \(\boldsymbol{\theta}^*\) is overly specific, making it indistinguishable from discriminative training and causing the parameterization to be redundant. For the NP-hard problems addressed in this paper, efficiently obtaining a large amount of ground-truth samples for discriminative supervised training is not feasible. In contrast, suboptimal solutions are significantly easier to obtain, which makes the solution distribution learning with the generative diffusion model possible.

Let $(\mathbf{x},\mathbf{y}')$ be a suboptimal solution for a given $\mathbf{x}$, and $(\mathbf{x},\boldsymbol{\theta}')$ be the parametrization. To avoid verbosity, we will not introduce the definition of the gap between suboptimal solutions and the optimal solution here and leave it in Sec. \ref{sec_converge}. We know that suboptimal parameterizations are not unique\footnote{For example, taking $\boldsymbol{\theta}'_1=(0.9,0.1),\boldsymbol{\theta}'_2=(0.95,0.05)$ for $\boldsymbol{\theta}^*=(1,0)$, where repeated sampling of $p_{\boldsymbol{\theta}'_1}(\mathbf{y}|\mathbf{x})$ or $p_{\boldsymbol{\theta}'_2}(\mathbf{y}|\mathbf{x})$ can easily hit an optimal solution.}. Therefore, let $Y'_{\mathbf{x}}$ be the set of suboptimal solutions for $\mathbf{x}$, and $\boldsymbol{\Theta}_{\mathbf{x}}$ be the set of suboptimal parameterizations which contains $\boldsymbol{\theta}^*$.

For the continuous solution space, it is impossible to make the calculated solution completely equal to $\mathbf{y}^*$ because of the infinite precision. As a discussion example, we parametrize $\mathbf{y}^*$ using sharp Laplace distribution, making the probability of the optimal solution approximate to 1. For example, for $y^*_i$, $\theta^*_i=y^*_i$ for ${\rm Laplace}(\theta^*_i,{\rm scale\_factor})$, smaller ${\rm scale\_factor}$ means sharper probability density function and the probability of $\mathbf{y}^*$ is the largest and tends to 1. Thus, we can obtain the continuous solution space's $Y'_{\mathbf{x}}$ and $\boldsymbol{\Theta}_{\mathbf{x}}$. 

\begin{figure*}[t]
\centering
\centerline{\includegraphics[width=5.57in]{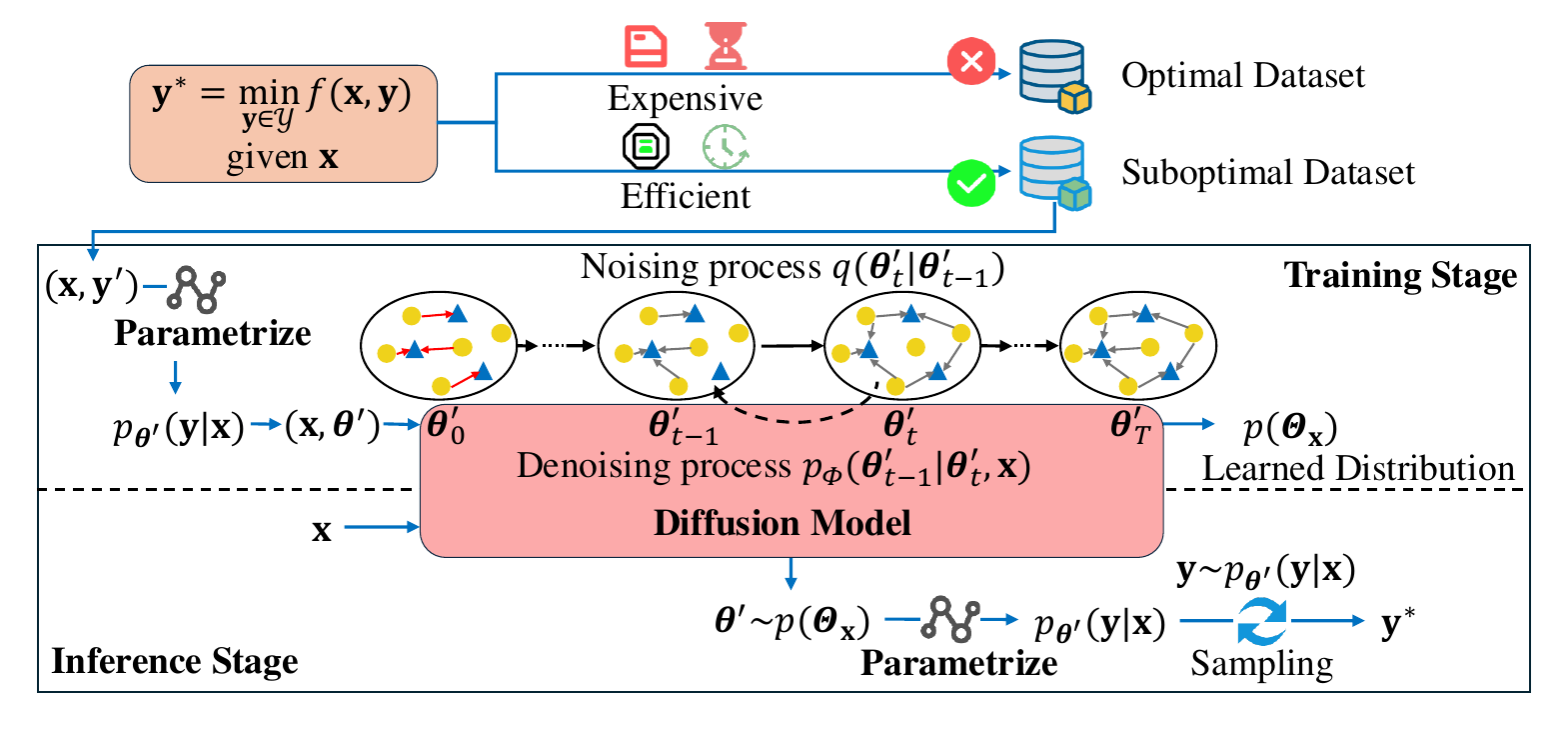}}
\setlength{\abovecaptionskip}{-0.1cm}
\caption{Framework for training the diffusion generative model with suboptimal dataset to achieve the optimal solution generation.}
\label{fig_sol_gen}
\vspace{-0.45cm}
\end{figure*}

{Consequently, when given only a dataset of suboptimal solutions, learning the parametrization of the solution space with parallel sampling can lead to the optimal solution.} Specifically, the model aims to learn $p(\boldsymbol{\Theta}_{\mathbf{x}})$ to obtain $\boldsymbol{\theta}'\!\sim\! p(\boldsymbol{\Theta}_{\mathbf{x}})$ and sample feasible solutions $\mathbf{y}\sim p_{\boldsymbol{\theta}'}(\mathbf{y}|\mathbf{x})$ to achieve the optimal solution, as shown in Fig. \ref{fig_sol_gen}. 

\subsection{Convergence to the Optimal Solution}\label{sec_converge}
In the previous section, we successfully transformed the goal into learning and sampling $p(\boldsymbol{\Theta}_{\mathbf{x}})$. In this section, we discuss the quality requirements of suboptimal solutions and the generation convergence to optimal solutions. 

\subsubsection{Gap between suboptimal and optimal parameterizations}
Let $(\mathbf{x},\boldsymbol{\theta}^*),\mathbf{x}\in\mathcal{X}$ and $(\mathbf{x},\boldsymbol{\theta}'),\boldsymbol{\theta}'\sim p(\boldsymbol{\Theta}_{\mathbf{x}})$. Although it is not possible to precisely quantify the magnitude of suboptimal perturbations, for the sake of initial discussion, we assume that the perturbations caused by suboptimal solutions are smoothly random and the mean $\mathbb{E}(\boldsymbol{\Theta}_{\mathbf{x}})=\boldsymbol{\mu}=\boldsymbol{\theta}^*$ holds. With $\mathbb{V}(\boldsymbol{\Theta}_{\mathbf{x}})=\boldsymbol{\sigma}^2$ as the variance of $p(\boldsymbol{\Theta}_{\mathbf{x}})$, we can obtain Chebyshev's inequality \cite{saw1984chebyshev} as below
\begin{equation}\label{eq_Chebyshev}
p\{|\boldsymbol{\theta}'-\boldsymbol{\theta}^*|\ge \boldsymbol{\epsilon}\}\leq\frac{\boldsymbol{\sigma}^2}{\boldsymbol{\epsilon}^2},\ \forall\boldsymbol{\epsilon}>0.
\end{equation} \vspace{-2mm}

Eq. (\ref{eq_Chebyshev}) presents the impact of the model solution generation variance and the quality of suboptimal solutions on the gap between the learned suboptimal and optimal parameterizations. By controlling the variance of generation in Eq. (\ref{eq_Chebyshev}), the probability that the parametrized $\boldsymbol{\theta}'$ of the model output falls outside the $\boldsymbol{\epsilon}$-neighborhood of $\boldsymbol{\theta}^*$ can be small. However, since the generation variance of the diffusion model cannot be theoretically quantified, we conduct simulations and analysis in the experimental section to explore this further. 

\subsubsection{Expectation of hitting $\mathbf{y}^*$}
For the $N$-dimensional binary classification problem in the discrete solution space, we have $|\theta_i-\theta^*_i|\!=\!\boldsymbol{\epsilon},\boldsymbol{\theta}\in\dot{U}(\boldsymbol{\theta}^*,\boldsymbol{\epsilon}),i\in\{1,...,N\}$, where $\dot{U}(\boldsymbol{\theta}^*,\boldsymbol{\epsilon})$ represents the center $\boldsymbol{\theta}^*$ and the neighborhood with radius $\boldsymbol{\epsilon}$. With $\theta^*_i=1$ generally, the probability of hitting $\mathbf{y}^*$ in a single sampling is $(1-\boldsymbol{\epsilon})^N$ with independent variables in each dimension. The probability of hitting $\mathbf{y}^*$ at least once in $n$ samplings can be calculated as $1-[1-(1-\boldsymbol{\epsilon})^N]^n$, which is finite. 

For the $N$-dimensional regression problem in the continuous solution space, let $\boldsymbol{\theta}\in\dot{U}(\boldsymbol{\theta}^*,\boldsymbol{\epsilon})$ have mean $\boldsymbol{\mu}+\boldsymbol{\epsilon}$ and variance $\boldsymbol{\gamma}^2$. Keeping the implication of the mean consistent, the parametrized distribution can be independent of the diffusion model sampling. In our discussion, we use Gaussian distribution $\mathcal{N}(\boldsymbol{\mu}+\boldsymbol{\epsilon},\boldsymbol{\gamma}^2)$ for feasible solution sampling. Take the tolerable gap as $\delta>0$, the probability of falling $[y^*_i-\delta,y^*_i+\delta]$ in a single sampling is 
\begin{equation}\label{eq_continuous_int}
\begin{aligned}
    P_i&=P(\mu_i-\delta\leq y_i\leq\mu_i+\delta) \\
    &=\int^{\mu_i+\delta}_{\mu_i-\delta} (1/\sqrt{2\pi\gamma^2_i}){\rm exp}(-\frac{(y-(\mu_i+\boldsymbol{\epsilon}))^2}{2\gamma^2_i})\ dy,
\end{aligned}
\end{equation}
where $y_i\sim\mathcal{N}(\mu_i+\boldsymbol{\epsilon},\gamma^2_i),\mu_i=y^*_i$. 

Also with independent variables in each dimension, the probability of hitting $\mathbf{y}^*$ at least once in $n$ samplings can be calculated as $1-(1-\prod^N_{i=1}P_i)^n$, which is finite. 

Given an $\boldsymbol{\epsilon}$-neighborhood of $\boldsymbol{\theta}^*$ as $\dot{U}(\boldsymbol{\theta}^*,\boldsymbol{\epsilon})$, and $\boldsymbol{\theta}$ is taken from $\dot{U}(\boldsymbol{\theta}^*,\boldsymbol{\epsilon})$. {Let $q(\boldsymbol{\theta}^n)$ be the probability of hitting $\mathbf{y}^*$ at least once in $n$ samples $\mathbf{y}\sim p_{\boldsymbol{\theta}}(\mathbf{y}|\mathbf{x})$, if $\boldsymbol{\epsilon}$ in Eq. (\ref{eq_Chebyshev}) makes the gap between $\boldsymbol{\theta}'$ and $\boldsymbol{\theta}^*$ sufficiently small, a finite number of solution sampling $n$ can make the expectation $\mathbb{E}(q(\boldsymbol{\theta}^n))\approx 1$.} %Therefore, the diffusion model is capable of learning a solution distribution $\boldsymbol{\theta}'$ that is sufficiently close to the optimal parametrization $\boldsymbol{\theta}^*$. 

\subsubsection{The lower bound on the number of samples}
To demonstrate the impact of $N$ and $n$ on the expectation of hitting $\mathbf{y}^*$, we provide Fig. \ref{fig_empirical_ver} and give the detailed calculation settings in Sec. \ref{sec_verify_y_theory}. 
Based on the above discussion, we can give a lower bound on $n$ that makes $\mathbb{E}(q(\boldsymbol{\theta}^n))\approx 1$ hold. By simplifying $p_{\boldsymbol{\epsilon}}=p\{|\boldsymbol{\theta}'-\boldsymbol{\theta}^*|\ge \boldsymbol{\epsilon}\}$ and letting $\mathbf{a}>\mathbf{0}$ and $\mathbf{a}\to \mathbf{0}$, we can multiply both sides of Eq. (\ref{eq_Chebyshev}) by $1-[1-(1-\boldsymbol{\epsilon})^N]^n$ to obtain 
\begin{equation}\label{eq_lower_Chebyshev}
\begin{split}
    (1-p_{\boldsymbol{\epsilon}})[1-(1-(1-\boldsymbol{\epsilon})^N)^n]>\ \ \ \ \ \ \ \ \ \ \ \ \ \ \ \ \ \ \ \ \ \ \ \ \\
    \ \ \ (1-\frac{\boldsymbol{\sigma}^2}{\boldsymbol{\epsilon}^2})[1-(1-(1-\boldsymbol{\epsilon})^N)^n]+\mathbf{a},\ \forall\boldsymbol{\epsilon}>0,
\end{split}
\end{equation}
where $1-[1-(1-\boldsymbol{\epsilon})^N]^n>0$ does not change the sign of the inequality, and the left side represents the probability that $\boldsymbol{\theta}\in\dot{U}(\boldsymbol{\theta}^*,\boldsymbol{\epsilon})$ and $\mathbb{E}(q(\boldsymbol{\theta}^n))\approx 1$ occur simultaneously. For a given $\mathbf{a}$, $p(\boldsymbol{\Theta}_{\mathbf{x}})$ in our problem is almost impossible to be the three-point distribution required by the equality condition in Chebyshev's inequality \cite{saw1984chebyshev}, so the sign `$\ge$' can be directly replaced by `$>$', thus an invalid inequality will not appear because $\mathbf{a}\to \mathbf{0}$. 

With the Eq. (\ref{eq_Chebyshev}) holds, we observe that $p_{\boldsymbol{\epsilon}}\leq\frac{\boldsymbol{\sigma}^2}{\boldsymbol{\epsilon}^2}$. After transposition and simplification, we can get the lower bound on $n$ for the discrete solution space:
\begin{equation}\label{eq_dis_lower_bound}
    n > \frac{{\rm ln}(\frac{\mathbf{a}}{p_{\boldsymbol{\epsilon}}-\boldsymbol{\sigma}^2/\boldsymbol{\epsilon}^2}+1)}{{\rm ln}(1-(1-\boldsymbol{\epsilon})^N)},\ \forall\boldsymbol{\epsilon}>0.
\end{equation}
For a continuous solution space, we use $1-(1-\prod^N_{i=1}P_i)^n$ to replace $1-(1-(1-\boldsymbol{\epsilon})^N)^n$ and the lower bound can be calculated as 
\begin{equation}\label{eq_con_lower_bound}
    n>\frac{{\rm ln}(\frac{\mathbf{a}}{p_{\boldsymbol{\epsilon}}-\boldsymbol{\sigma}^2/\boldsymbol{\epsilon}^2}+1)}{{\rm ln}(1-\prod^N_{i=1}P_i)},\ \forall\boldsymbol{\epsilon}>0.
\end{equation}

To make $\mathbb{E}(q(\boldsymbol{\theta}^n))\approx 1$ hold, the lower bound of the number of solution sampling $n$ is given by (\ref{eq_dis_lower_bound}) and (\ref{eq_con_lower_bound}), respectively. {Therefore, the diffusion model with solution sampling $n$ is capable of learning a solution distribution $\boldsymbol{\theta}'$, which is sufficiently close to the optimal parametrization $\boldsymbol{\theta}^*$.} To demonstrate the impact of $N$ on the lower bound on $n$ that supports $\mathbb{E}(q(\boldsymbol{\theta}^n))\approx 1$, we illustrate it in Fig. \ref{fig_empirical_ver} and the detailed settings in Sec. \ref{sec_verify_y_theory}.

%\begin{observation}
%The diffusion model is capable of learning a solution distribution $\boldsymbol{\theta}'$ that is sufficiently close to the optimal parametrization $\boldsymbol{\theta}^*$. Furthermore, within a finite number of samples $n$, the expectation of hitting the optimal solution at least once can approach $1$.
%\end{observation} %Exploratively, we provide a lower bound expression (Eq. (\ref{eq_dis_lower_bound}), Eq. (\ref{eq_con_lower_bound})) for the number of samples $n$ that supports this expectation approaching $1$. These equations are further studied by experiments and model results, thereby providing valuable basis and reference conditions. 

\section{The Proposed GDSG Algorithm} 
With the solution distribution parametrization provided, the core theoretical concept in Sec. \ref{sec_method_def} is to establish a model capable of generating $\boldsymbol{\theta}'\sim p(\boldsymbol{\Theta}_{\mathbf{x}})$ for a given $\mathbf{x}$. Generative models aim to model the joint distribution of inputs and outputs, $P(\boldsymbol{\Theta}_{\mathbf{x}}, \mathbf{x})$, in contrast to discriminative models, which learn the conditional probability distribution $P(\boldsymbol{\Theta}_{\mathbf{x}}|\mathbf{x})$. Consequently, the generative model uses the implicitly learned $P(\boldsymbol{\Theta}_{\mathbf{x}}|\mathbf{x})$ as a guide to generate $\boldsymbol{\theta}'\sim p(\boldsymbol{\Theta}_{\mathbf{x}})$ in conditional generation \cite{ho2022classifier}, thereby enhancing the likelihood of producing high-quality outputs. To address the graph optimization problem outlined in Sec. \ref{sec_problem} and implement the approach outlined in Sec. \ref{sec_method_def}, we propose a graph diffusion generative model for solution generation. \textcolor{black}{The concepts and analysis in Section \ref{sec_method_def} can be applied to various generative models. However, due to the diffusion model's capability to accommodate different data types and its versatility across various model scales, we focus on GDM in this work.}

\subsection{Graph Diffusion for Solution Generation}
From the perspective of variational inference, the general diffusion framework consists of a forward-noising process and a learnable reverse denoising Markov process \cite{ho2020denoising,kingma2024understanding,kingma2021variational}. 

As described in Sec. \ref{sec_problem}, for an input $\mathcal{G}(\mathcal{V},\mathcal{E})$, the optimal solution to the multi-server multi-user computation offloading problem can be expressed as $\{\mathcal{D}^*,\mathcal{A}^*\}$. Next, we introduce the diffusion generation of $\mathcal{D}^*$ and $\mathcal{A}^*$ in the following two aspects: discrete solutions and continuous solutions generation. 

\subsubsection{Discrete Solution Diffusion} For discrete solution $\mathbf{y} \in \mathcal{Y}$, it is parametrized as a one-hot vector that $\mathbf{y} \in \{0,1\}^{L \times 2}$. Given an instance $(\mathcal{G}, \mathbf{y})$, diffusion takes $\mathbf{y}$ as the initial solution $\mathbf{y}_0$ to perform the noising process, also known as the diffusion process, gradually introducing noise to obtain a series of variables $\mathbf{y}_{1:T} = \mathbf{y}_1, \ldots, \mathbf{y}_T$ with $T$ as the diffusion steps. The specific noising process can be formalized as $q(\mathbf{y}_{1:T}|\mathbf{y}_0)=\prod^T_{t=1}q(\mathbf{y}_t|\mathbf{y}_{t-1})$. The model learns the noise distribution from this process, enabling it to denoise from a given initial noise distribution to a target distribution. The denoising process actually starts from $\mathbf{y}_T$ and performs adjacent step denoising from $\mathbf{y}_t$ to $\mathbf{y}_{t-1}$ with $\mathcal{G}$ as the condition. The denoising process can be expressed as $p_{\Phi}(\mathbf{y}_{0:T}|\mathcal{G})=p(\mathbf{y}_T)\prod^T_{t=1}p_{\Phi}(\mathbf{y}_{t-1}|\mathbf{y}_t,\mathcal{G})$, where $\Phi$ represents the model parameters. The ideal result of diffusion model training is $p_{\Phi}(\mathbf{y}|\mathcal{G})=q(\mathbf{y}|\mathcal{G})$, and to fit the original data distribution, the variational upper bound of the negative log-likelihood is generally used as the loss objective 
\begin{align}\label{eq_loss_elbo}
    &\mathcal{L}=\mathbb{E}_q\big[-\mathrm{log}p_{\Phi}(\mathbf{y}_0|\mathbf{y}_1,\mathcal{G}) \notag \\
    &+\prod_{t>1}D_{KL}[q(\mathbf{y}_{t-1}|\mathbf{y}_t,\mathbf{y}_0)\ ||\ p_{\Phi}(\mathbf{y}_{t-1}|\mathbf{y}_{t},\mathcal{G})]\big] + C,
\end{align}
where $D_{KL}$ represents Kullback–Leibler divergence, a measure of the difference between two distributions, and $C$ represents a constant. 

In the forward noising process, $\mathbf{y}_t$ is multiplied by $\mathbf{Q}_t$ to obtain $\mathbf{y}_{t+1}$, where $\mathbf{Q}_t=\begin{bmatrix}
    (1-\beta_t) & \beta_t \\ 
    \beta_t & (1-\beta_t)
\end{bmatrix}$ is the transition probability matrix \cite{austin2021structured,hoogeboom2021argmax}. The operation $\mathbf{y}\mathbf{Q}$ converts and adds noise to the one-hot vector of each element $y_i\in\{0,1\}^2$, with $\beta_t$ controlling the noise scale as the corruption ratio. Specifically, $\prod^T_{t=1}(1-\beta_t)\approx 0$ ensures that $\mathbf{y}_T\sim \mathrm{Uniform}(\cdot)$. Let $\overline{\mathbf{Q}}_t=\mathbf{Q}_1\mathbf{Q}_2\ldots\mathbf{Q}_t$, the single-step and $t$-step marginals of the noising process are formulated as 
\begin{align}
    q(\mathbf{y}_t|\mathbf{y}_{t-1})&=\mathrm{Cat}(\mathbf{y}_t;\mathbf{p}=\mathbf{y}_{t-1}\mathbf{Q})\ \ \ \mathrm{and} \notag \\
    q(\mathbf{y}_t|\mathbf{y}_0)&=\mathrm{Cat}(\mathbf{y}_t;\mathbf{p}=\mathbf{y}_0\overline{\mathbf{Q}}_t),
\end{align}
where $\mathrm{Cat}(\cdot)$ represents a categorical distribution, which can be any discrete distribution. Here, in line with experience setting \cite{austin2021structured}, a Bernoulli distribution is applied. 

According to Bayes' theorem, the posterior can be calculated as 
\begin{align}\label{eq_discrete_posterior}
    &q(\mathbf{y}_{t-1}|\mathbf{y}_t,\mathbf{y}_0)=\frac{q(\mathbf{y}_t|\mathbf{y}_{t-1},\mathbf{y}_0)q(\mathbf{y}_{t-1}|\mathbf{y}_0)}{q(\mathbf{y}_t|\mathbf{y}_0)} \notag \\
    &=\mathrm{Cat}(\mathbf{y}_{t-1};\mathbf{p}=\frac{\mathbf{y}_t\mathbf{Q}^T_t\odot\mathbf{y}_0\overline{\mathbf{Q}}_{t-1}}{\mathbf{y}_0\overline{\mathbf{Q}}_t \mathbf{y}^T_t}),
\end{align}
where $\odot$ denotes element-wise multiplication. 

The model is trained to predict the distribution $p_{\Phi}(\tilde{\mathbf{y}}_0|\mathbf{y}_t,\mathcal{G})$, allowing the denoising process to be obtained by substituting the $\mathbf{y}_0$ in Eq. (\ref{eq_discrete_posterior}) with the estimated $\tilde{\mathbf{y}}_0$
\begin{equation}\label{eq_dis_denoise_pos}
    p_{\Phi}(\mathbf{y}_{t-1}|\mathbf{y}_t)\propto \sum_{\tilde{\mathbf{y}}_0}q(\mathbf{y}_{t-1}|\mathbf{y}_t,\tilde{\mathbf{y}}_0)p_{\Phi}(\tilde{\mathbf{y}}_0|\mathbf{y}_t,\mathcal{G}).
\end{equation}

\subsubsection{Continuous Solution Diffusion} For the continuous solution $\mathbf{y} \in [0,1]^L$, continuous diffusion models \cite{ho2020denoising,ho2022classifier,dhariwal2021diffusion} can be directly applied without parametrization. Unlike previous studies \cite{sun2023difusco,li2024distribution} rescaling discrete data in continuous diffusion for discrete data generation \cite{chen2022analog}, our multi-task generation approach uses continuous diffusion for continuous solutions. 

The noising and denoising process formulas for continuous diffusion are essentially the same as those for discrete diffusion, and the loss objective is similar to Eq. (\ref{eq_loss_elbo}). The primary difference is that the transition probability in continuous diffusion is Gaussian rather than categorical. Therefore, let $\beta_t$ be the corruption ratio, the single-step and t-step marginals of the noising process for continuous diffusion can be formulated as 
\begin{align}
    q(\mathbf{y}_t|\mathbf{y}_{t-1})&=\mathcal{N}(\mathbf{y}_t;\sqrt{1-\beta_t}\mathbf{y}_{t-1},\beta_t\mathbf{I})\ \  \notag \\
    q(\mathbf{y}_t|\mathbf{y}_0)&=\mathcal{N}(\mathbf{y}_t;\sqrt{\overline{(1-\beta_t)}}\mathbf{y}_0,(1-\overline{(1-\beta_t)})\mathbf{I}),
\end{align}
where $\overline{(1-\beta_t)}=\prod^t_{t=1}(1-\beta_t)$, and $\prod^T_{t=1}(1-\beta_t)\approx 0$ ensures that $\mathbf{y}_T\sim \mathcal{N}(\cdot)$. Similar to Eq. (\ref{eq_discrete_posterior}), the posterior can be calculated as 
\begin{equation}\label{eq_continuous_posterior}
    q(\mathbf{y}_{t-1}|\mathbf{y}_t,\mathbf{y}_0)=\frac{q(\mathbf{y}_t|\mathbf{y}_{t-1},\mathbf{y}_0)q(\mathbf{y}_{t-1}|\mathbf{y}_0)}{q(\mathbf{y}_t|\mathbf{y}_0)}.
\end{equation}

The model is trained to predict the Gaussian noise $\tilde{\mathbf{\boldsymbol{\epsilon}}}_t=f_{\Phi}(\mathbf{y}_t,t,\mathcal{G})$. Given that $\tilde{\mathbf{\boldsymbol{\epsilon}}}_t=(\mathbf{y}_t-\sqrt{\overline{(1-\beta_t)}}\mathbf{y}_0)/\sqrt{1-\overline{(1-\beta_t)}}$, the denoising process can be obtained by substituting the $\mathbf{y}_0$ in Eq. (\ref{eq_continuous_posterior}) with an estimation 
\begin{equation}\label{eq_con_denoise_pos}
    p_{\Phi}(\mathbf{y}_{t-1}|\mathbf{y}_t)\propto q(\mathbf{y}_{t-1}|\mathbf{y}_t,\frac{\mathbf{y}_t-\sqrt{1-\overline{(1-\beta_t)}}f_{\Phi}(\mathbf{y}_t,t,\mathcal{G})}{\sqrt{\overline{(1-\beta_t)}}}).
\end{equation}

\subsection{GNN-based Diffusion Model}\label{sec_gnn}
In Sec. \ref{sec_problem}, we model the computation offloading in the MEC network as a graph optimization problem to enhance the model's versatility in network optimization. The optimization variables are represented as discrete solutions for edge selection and continuous solutions for edge weights, corresponding to offloading decisions and computational resource allocation ratios. For the model prediction objectives in Eq. (\ref{eq_dis_denoise_pos}) and Eq. (\ref{eq_con_denoise_pos}), theoretically, any neural network structure can be employed. We use a convolutional GNN with a specified edge gating mechanism, which is designed to mitigate the effects of padding edges \cite{joshi2022learning,dwivedi2023benchmarking}. We employ a multi-task approach to expand the model, supporting the prediction objectives in Eq. (\ref{eq_dis_denoise_pos}) and Eq. (\ref{eq_con_denoise_pos}). 

The core operation of the GNN layer involves transferring or aggregating edge and node feature information locally or globally \cite{guo2022systematic}, with various options for aggregation operation. Additionally, the edge gating mechanism controls the nonlinear activation of edge features before aggregation using off-the-shelf functions. Specifically, the GNN takes the current step $t$, the feature graph $\mathcal{G}$, and the current solution with the same data structure as $\mathcal{G}$ as input. For known node and edge feature dimensions, the model embeds them into a latent space of preset dimensions (usually high dimensions) and performs recursive information transfer and aggregation in this latent space. Information extraction and fusion in the latent space can be considered the source of the GNN's permutation-invariance.

However, we notice that the existing works \cite{sun2023difusco,li2024distribution} using convolutional GNN only consider fully connected input graphs for the TSP and MIS problems, whereas most graph optimization problems are not fully connected. In dense adjacency matrix and sparse edge list graph data structures, some edges may not exist, which we refer to as padding edges. The feature encoding of these padding edges can have unpredictable (and often adverse) effects on the iteration of the GNN in the latent space. This issue is not well addressed by previous works. 

We investigate this problem during our experiments, as detailed in Sec. \ref{sec_vs_gnndi}. Specifically, the feature embedding of padding edges is non-zero. The input graphs of different sizes may have varying proportions of padding edges, and their influence is difficult to generalize or eliminate through simple linear or nonlinear gating. Changing the aggregation method (e.g., sum, mean, max) offers little help. Moreover, we considered setting the original features of padding edges to zero and setting the bias of the edge feature embedding layer to zero simultaneously. However, this approach could adversely affect the embedding of normal edges and might still not completely distinguish padding edges from normal edges in some cases. 

To address this, we design a gating mechanism that can mitigate the effects of padding edges, inspired by the spirit that the generalization performance of GNNs benefits from explicit redesign \cite{joshi2022learning,guo2022systematic}. We add a one-dimensional 0-1 variable to the edge feature to indicate whether an edge is a padding edge. This variable multiplies the original gating result, eliminating the influence of meaningless hidden features from padding edges. Notably, the settings of node and edge features can be flexibly adjusted to support various MEC tasks.

 {The proposed GDSG method can be extended to multi-task capabilities and improve generalization for both fully connected and non-fully connected graphs.} Additionally, as an anisotropic GNN handling directionally dependent features, our model is not limited to the directed graph problem and can be also applied to undirected graphs. GNNs inherently offer versatility, allowing a single trained model to handle graphs of different scales for the same problem. 

\subsection{Multi-Task Generation and Denoising Schedule}
%Our model performs multi-task generation, denoising both the target discrete solution and the continuous solution. In multi-task learning, negative competition between tasks can lead to poor convergence. However, we find that the discrete and continuous diffusion tasks largely satisfy the metric of inter-task gradient direction orthogonality, a key goal in specialized multi-task learning research \cite{liu2021towards,yu2020gradient,NEURIPS2020_sign}. Our experiments reveal that the gradient vectors calculated on the same layer parameters for the two tasks are almost orthogonal, which is detailed in Sec. \ref{sec_orth}. A comparison using the same neural network structure for discriminative training confirms that this orthogonality is not inherent to the network architecture. We believe this characteristic is unique to the training loss objectives of discrete and continuous diffusion, a phenomenon not yet reflected in existing multi-task learning research on diffusion models \cite{go2024addressing,ye2024diffusionmtl,NEURIPS2023_effective}. 

Our model is designed for multi-task generation, specifically focusing on denoising both the target discrete solution and the continuous solution. In the area of multi-task learning \cite{liu2021towards,yu2020gradient}, it is recognized that negative competition between tasks can impede convergence. \textcolor{black}{During training, the gradient directions of different task losses are often misaligned when they share the same parameters. Simply combining these non-orthogonal gradients may lead to suboptimal training results, sometimes causing worse performance on individual tasks compared to single-task training. Our approach, GDSG, incorporates both continuous and discrete diffusion, with the two tasks sharing all parameters except for the output heads. Therefore, we analyze this interaction carefully to mitigate potential degradation in training performance.}

However, we discover that the discrete and continuous diffusion tasks primarily align with the principle of inter-task gradient direction orthogonality, which is a significant goal in specialized multi-task learning research \cite{liu2021towards,yu2020gradient,NEURIPS2020_sign}. We find that the gradient vectors calculated for these two tasks, using the same layer parameters, are nearly orthogonal. This finding is further elaborated in Section \ref{sec_orth}. A comparative study using the same neural network architecture for discriminative training indicates that this orthogonality does not stem from the network architecture itself. We posit that this characteristic is intrinsic to the training loss objectives for discrete and continuous diffusion, representing a phenomenon not yet captured in the current multi-task learning research focused on diffusion models~\cite{go2024addressing,ye2024diffusionmtl,NEURIPS2023_effective}. \textcolor{black}{We observe that task orthogonality is not only an impressive side effect of our method, it also demonstrates the GDM's great potential for scalability in tasks.}

Additionally, we employ DDIM to accelerate the denoising inference of the diffusion model~\cite{song2021denoising}. DDIM rewrites the denoising formula, leveraging the independence of marginal probabilities between two adjacent steps in the original process. This approach significantly reduces the number of inference steps while maintaining denoising performance. %For discrete diffusion, we use a similar method \cite{austin2021structured}. 

\section{Experiments}

To verify our proposed methods and models, we establish instances of the MSCO problem at various scales. We first provide targeted simulation validation for the approach outlined in Sec. \ref{sec_method_def}, and then we conduct a gradient analysis for multi-task orthogonality and compare GDSG with a discriminative model of the same neural network structure. Finally, we compare GDSG with multiple baselines in terms of cost metrics. 
\subsection{Experiments Settings}
\subsubsection{Datasets} \textcolor{black}{We tentatively tried using existing numerical solvers such as Gurobi and GEKKO to generate non-graph data for a single server, but the solution time for a single solve of a four-node MINLP problem reached several seconds. Moreover, it is difficult for Gurobi and GEKKO to handle graph structure data.}
As a result, we use a heuristic algorithm with polynomial time complexity to generate suboptimal datasets for the computation offloading problem in Sec. \ref{sec_problem}. The minimum cost maximum flow (MCMF) algorithm, applied to random heuristic initialization of edge weights, efficiently finds optimal edge selection results and provides high-quality suboptimal solutions (solving a single instance with $80$ nodes in $0.6$sec). 

\textcolor{black}{Specifically, MSCO can be formulated as a graph-level mixed-integer programming problem, where the discrete offloading decisions represent the data flow from users to servers. This formulation allows the problem to be solved using the MCMF algorithm. Initially, a heuristic is used to allocate computational resources and establish the continuous solution, i.e., 
$A_i = \frac{V_i}{\sum_{l,\Upsilon(E_l, U_j)\Omega(E_l, S_k)=1} V_j}, \forall i \in \{1, \dots, M\},\ k \in \{1, \dots, K\}$. By applying additional randomness, we can compute the costs for both local execution and offloading, which allows us to determine the outbound flow costs for users. To meet the integer cost requirement of the MCMF algorithm, we retain three decimal places in the real-valued costs and scale these values by $1000$. The MCMF algorithm then computes the minimum-cost flow while ensuring maximum flow throughput. Our topology design guarantees that the maximum flow in the MCMF formulation always matches the number of computational tasks. This effectively reduces the problem to a straightforward minimum-cost flow, optimizing the offloading decision with minimal computation cost. By repeatedly initializing the computational resource allocation randomly, solving the MCMF, and selecting the best solution, we efficiently generate high-quality suboptimal solution samples.} Since an optimal solver is not available, we use exhaustive methods to generate small-scale ground-truth datasets but the time complexity is very high (over $30$sec for a single instance of $80$ nodes). 

We generate $14$ training datasets with varying problem scales: $\rm lq3s6u$, $\rm lq3s8u$, $\rm lq4s10u$, $\rm lq4s12u$, $\rm lq7s24u$, $\rm lq7s27u$, $\rm lq10s31u$, $\rm lq10s36u$, $\rm lq20s61u$, $\rm lq20s68u$, $\rm gt3s6u$, $\rm gt3s8u$, $\rm gt4s10u$, and $\rm gt4s12u$. For example, ``$\rm lq3s6u$" represents a suboptimal dataset with $3$ servers and $6$ users at relatively low-quality, while ``$\rm gt3s6u$" represents the optimal dataset with $3$ servers and $6$ users at ground-truth. Our scale setting considers the radio coverage of edge servers, ensuring each user can only connect to a subset of servers. Ultra-sparse and ultra-dense settings lacking practical significance for computation offloading are not considered. Each suboptimal training dataset contains either $60000$ or $80000$ samples, while each optimal training dataset contains $2000$ samples due to the complexity. The testing sets consist of $10$ different problem scales: $\rm gt3s6u$, $\rm gt3s8u$, $\rm gt4s10u$, $\rm gt4s12u$, $\rm gt7s24u$, $\rm gt7s27u$, $\rm gt10s31u$, $\rm gt10s36u$, $\rm gt20s61u$, and $\rm gt20s68u$. We provide thousands of test instances for testing sets. None of the test samples appeared in the training set. 

\begin{figure*}
\flushleft
\begin{minipage}[t]{0.32\linewidth}
\centering
\setlength{\abovecaptionskip}{-0.1cm}
\includegraphics[width=2.3in, height=1.88in]{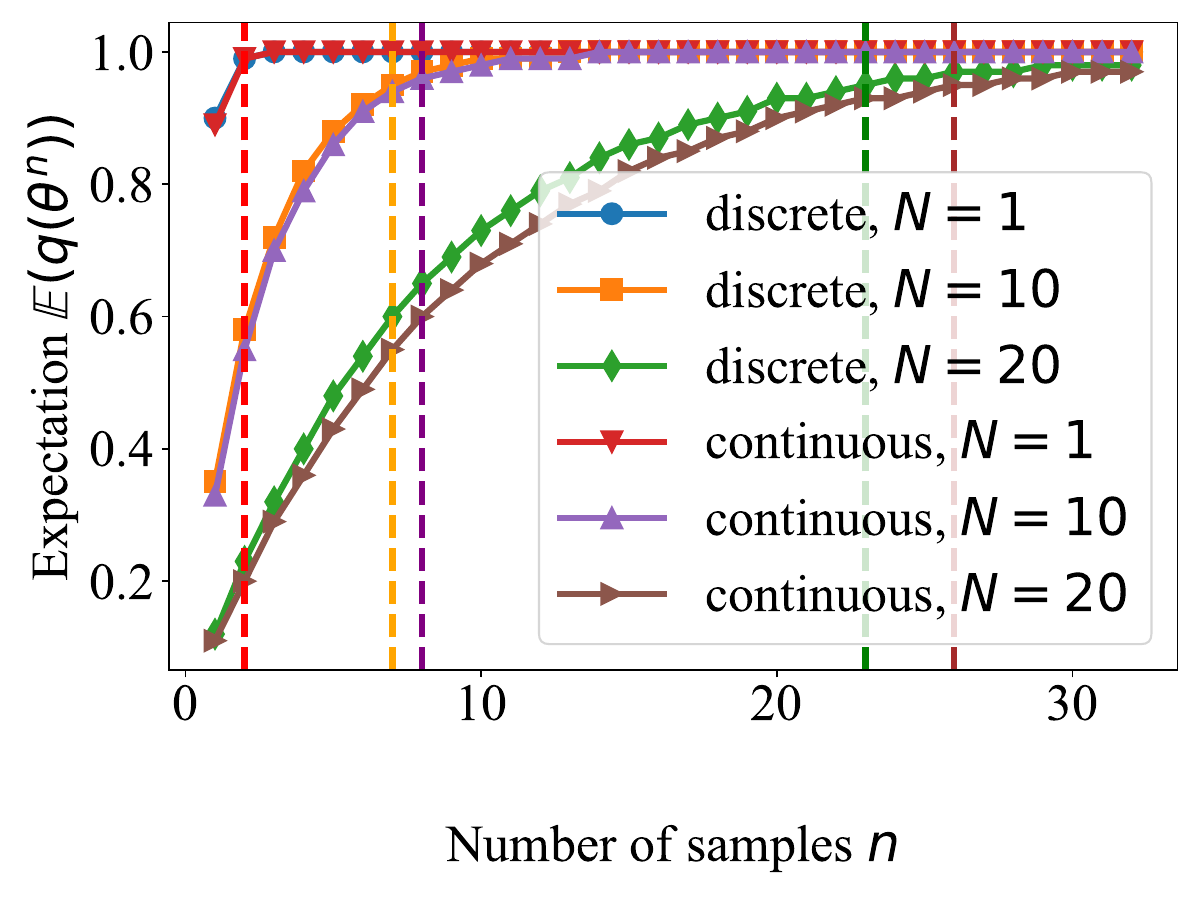}
\caption{The expectation of hitting $\mathbf{y}^*$ for different number of samples $n$ and variable dimension $N$. The dotted lines represent the lower bounds on $n$, where the red and black dotted lines coincide.}
% \vspace{-0.48cm}
\label{fig_empirical_ver}
\end{minipage}
\hskip 0.4cm
\begin{minipage}[t]{0.62\linewidth}
\setlength{\abovecaptionskip}{-0.1cm}
  \includegraphics[width=2.3in, height=1.88in]{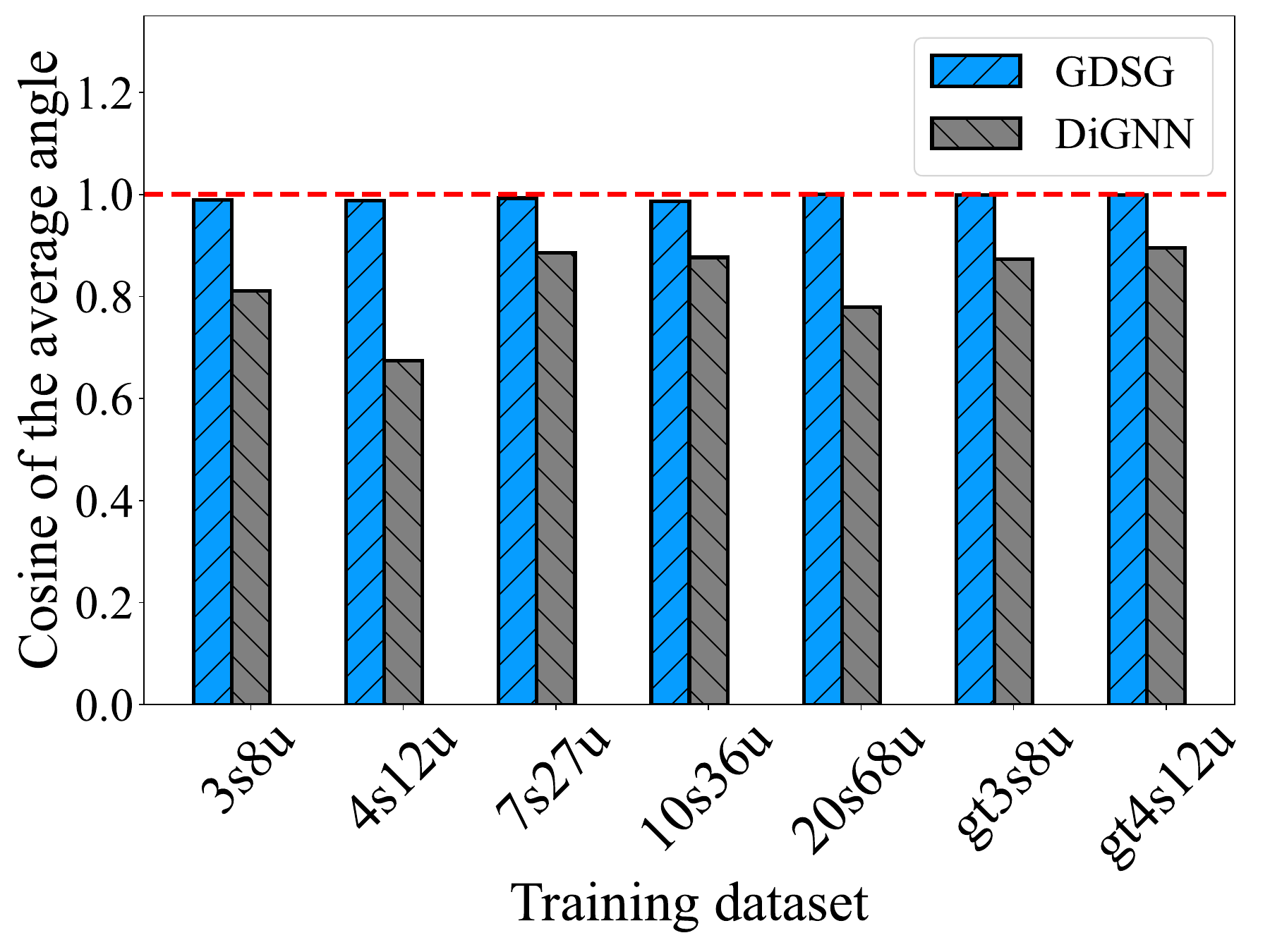}
  % \hspace{0.1cm}
  \includegraphics[width=2.3in, height=1.88in]{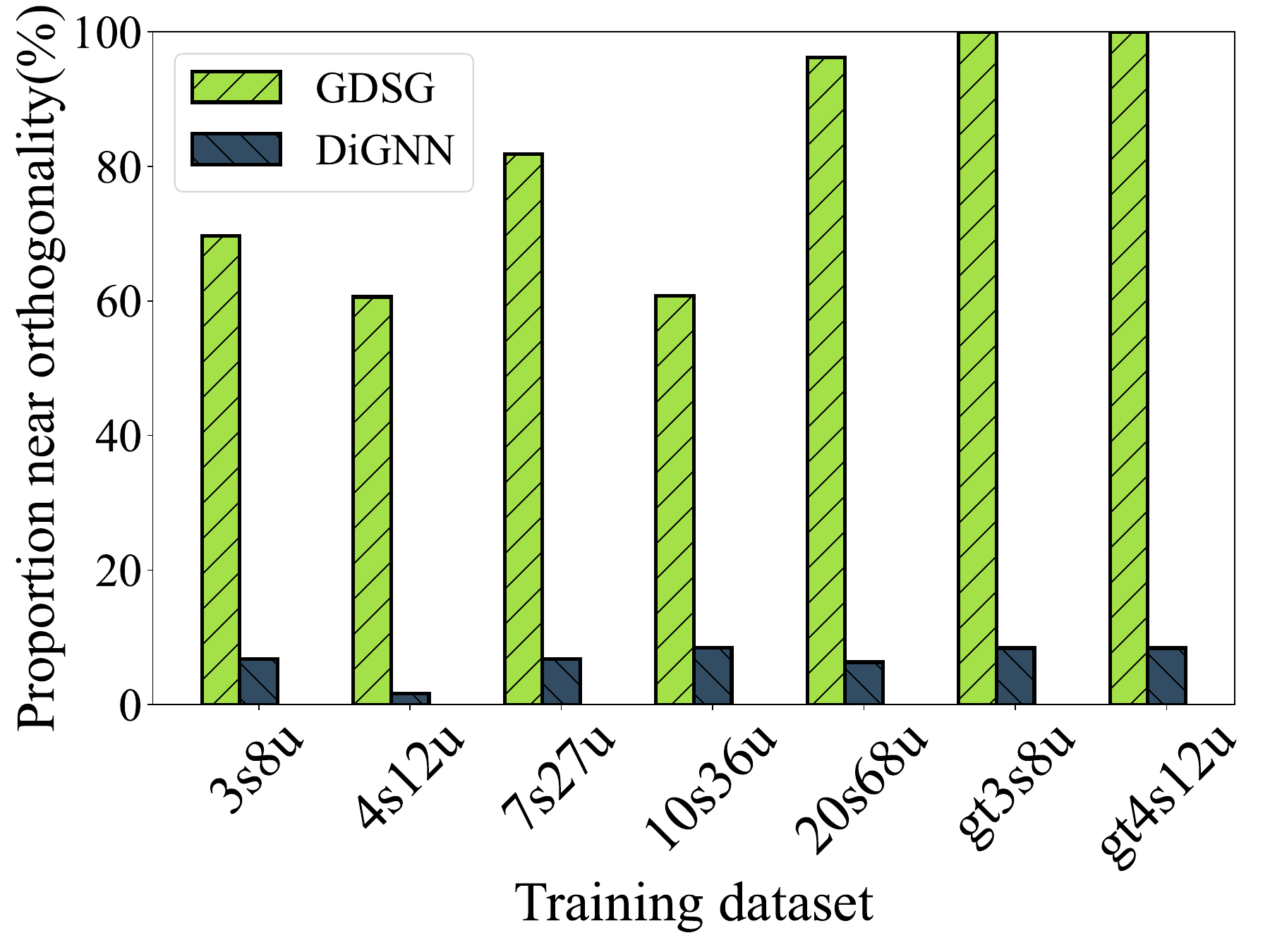}
\caption{Left: The cosine value of the overall average of all relevant parameters (The closer to 1, the better the orthogonality). 
Right: The proportion of the orthogonal vector for a given range in a training step (The closer to 100\%, the better the orthogonality).}
\label{fig_gradient_ana}
\end{minipage}
\vspace{-0.48cm}
\end{figure*}

\subsubsection{Model Settings} The main hyperparameters include $\rm diffusion\_steps$, $\rm GNN\_layer\_num$, $\rm hidden\_dim$, $\rm batch\_size$, $\rm epochs$, $\rm inference\_steps$, and $\rm parallel\_sampling\_number$. While the model can handle any graph scale after one training, we adjust settings for different training sets, detailed in our open-source code. Notably, GDSG uses very few diffusion (e.g., $200$) and inference steps (e.g., $5$), unlike the thousands of steps in other works \cite{ho2020denoising,sun2023difusco}. This efficiency is due to the lower-dimensional scale and higher certainty of our problem compared to image generation. 

\subsubsection{Metrics} We aim to minimize the weighted cost, so we use the  $\rm Exceed\_ratio$ of output solution cost to the testing label cost as the metric. On the optimal testing set, the $\rm Exceed\_ratio$ should be close to $1$, with smaller values being better. 

\subsubsection{Baselines} We use five baseline methods: the heuristic algorithm with the MCMF algorithm (HEU), which is the tool used to generate the suboptimal datasets for GDSG and other baselines, and the corresponding code has also been open-sourced; the discriminative GNN (DiGNN) with the same neural network structure as our GDSG; and MTFNN \cite{yang2020computation}, an off-the-shelf multi-task learning DNN method; DIFUSCO \cite{sun2023difusco} and T2T \cite{li2024distribution} are also included as latest known methods that use diffusion generation for combinatorial optimization. For all experiments, each hyperparameter for a given model and training set is fixed by default (details are in the open source). In the following sections, the neural network settings for GDSG and DiGNN are identical. GDSG shares the same diffusion-related hyperparameter settings as DIFUSCO and T2T. All dataset generation is done on an Intel i7-13700F, and all model training and testing are performed on an RTX 4090. 

\subsection{Empirical Verification for Theory}\label{sec_verify_y_theory}
\textcolor{black}{This experiment aims to clarify the theoretical concepts discussed in Section \ref{sec_converge} by demonstrating the expected probability of achieving an optimal solution through repeated sampling with specified hypothetical parameters. It also aims to illustrate the feasibility of learning the solution distribution necessary for reaching the optimal solution and analyzing key factors that influence the expected hit rate. These factors include the quality of suboptimal solutions within the training set and the dimensionality of the solution variables.}

Based on (Eq. (\ref{eq_dis_lower_bound}), Eq. (\ref{eq_con_lower_bound})), we demonstrate $\mathbb{E}(q(\boldsymbol{\theta}^n))$ for three cases in both discrete and continuous diffusion: $N=1, 10, 20$. We set the possible values for the unknown variables in Eq. ((\ref{eq_Chebyshev}), Eq. (\ref{eq_continuous_int})) under the assumptions in Sec. \ref{sec_converge}. For discrete solutions, we set $\boldsymbol{\epsilon}\!=\!0.1$. For continuous solutions, we uniformly set $y_i=1$, $\boldsymbol{\epsilon}\!=\!0.1$, $\boldsymbol{\gamma}_i\!=\!0.04$, and $\delta=0.1$. The expected value of $\mathbb{E}(q(\boldsymbol{\theta}^n))$ is then computed, resulting in Fig. \ref{fig_empirical_ver}. We use dotted lines of the same color as the curves to mark the corresponding lower bound on the number of samples $n$ in Eq. (\ref{eq_dis_lower_bound}) and Eq. (\ref{eq_con_lower_bound}). Here, we consider $\mathbb{E}(q(\boldsymbol{\theta}^n)) > 0.95$ as $\mathbb{E}(q(\boldsymbol{\theta}^n))\approx 1$. As $n$ increases, the expectation approaches $1$. Although some unknown variables are assigned assumed values, Fig. \ref{fig_empirical_ver} illustrates the convergence process of sampling from the high-quality solution distribution and the associated influencing factors. 

Additionally, for the unknown variables in Eq. (\ref{eq_Chebyshev}), which refer to the model's generation data variance, we use empirical calculations based on the existing model outputs. The trained model is used to repeat 1000 inferences on the same input and count the variance of the output $\boldsymbol{\theta}$s, which is on the order of $10^{-4}$. Therefore, $\frac{\boldsymbol{\sigma}^2}{\boldsymbol{\epsilon}^2}$ can be small, e.g., $\frac{\boldsymbol{\sigma}^2}{\boldsymbol{\epsilon}^2}=0.01\ if\ \boldsymbol{\sigma}^2=0.0001\ \boldsymbol{\epsilon}=0.1$. Although this is not a perfect theoretical conclusion, it demonstrates a small, even almost negligible, gap between the learned distribution and the optimal distribution. 

\subsection{Multi-Task Orthogonality}\label{sec_orth}
As we address discrete diffusion and continuous diffusion in a single model, we should ensure the gradient vectors of the two tasks are orthogonal so there is no negative competition between the two tasks during training \cite{liu2021towards,yu2020gradient}. The non-orthogonality of the gradient vectors of two tasks often leads to inferior convergence. Considering this challenge, we track the gradient directions of the related parameters for both tasks throughout the training process of our model and the baseline, DiGNN, with the same neural network structure. Specifically, we perform backpropagation of the loss for each task at each training iteration across multiple epochs.
Then, we extract a high-dimensional vector that captures the information from the parameter-updating gradient for each parameter of the representative network layers. \textcolor{black}{This includes the node feature embedding layer, the edge feature embedding layer, the three edge convolutional layers, and the learnable output normalization layer of the last graph convolutional module.} Finally, we compute the cosine of the angle between the update vectors of the two tasks to quantify their orthogonality. The results are shown in Fig. \ref{fig_gradient_ana}. 

It is observed that in GDSG, the two tasks show a high degree of orthogonality with cosine values approaching $1$ and orthogonal gradient proportions approaching $100\%$, which is in sharp contrast to the baseline. For DiGNN, it shows poor orthogonality with cosine values far away from 1 and orthogonal gradient proportions always less than $10\%$. Furthermore, GDSG has a small variance (the maximum magnitude is $4$) while DiGNN has a large variance (the minimum magnitude is $7$), which means that the gradient angles of GDSG are more concentrated in orthogonality. This advantage is benefit from the configuration of the loss function of diffusion. Therefore, the simultaneous training of the two tasks in a single GDSG model does not adversely affect each other, which avoids any degradation of model performance. %For this inspiring finding of multi-task learning that is beyond the scope of our study, we do not go into details with only empirical analysis here. 

\begin{figure}[t]
\centering
\setlength{\abovecaptionskip}{-0.1cm}
\centerline{\includegraphics[width=3.48in, height=2.76in]{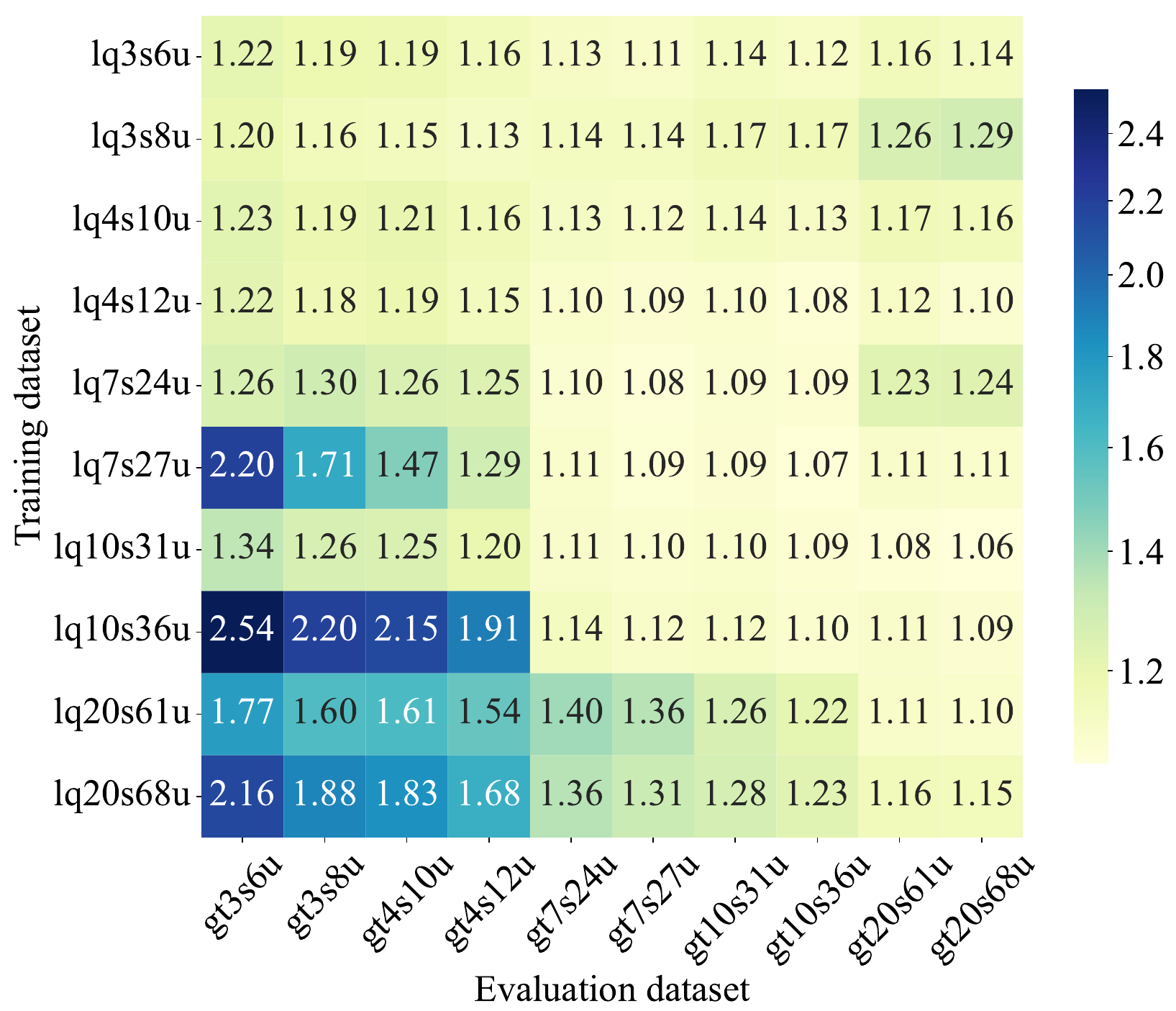}}
\caption{The performance Exceed\_ratio of GDSG (GNN padding edge not handled).}
\label{fig_GDSG_before_heatmap}
\vspace{-0.45cm}
\end{figure}

\begin{figure}[t]
\centering
\setlength{\abovecaptionskip}{-0.1cm}
\centerline{\includegraphics[width=3.48in, height=2.76in]{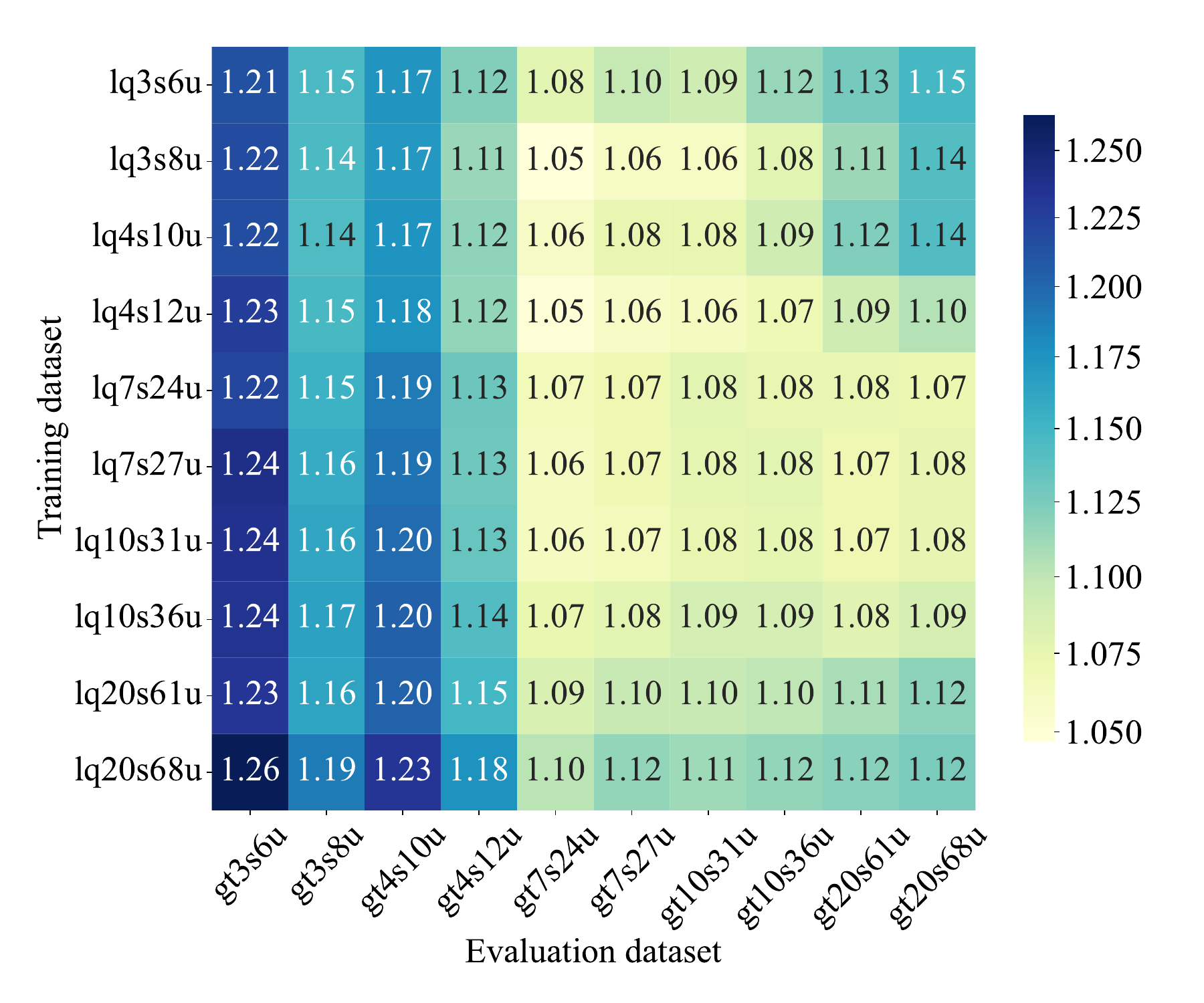}}
\caption{The performance Exceed\_ratio of GDSG (GNN padding edge handled).}
\label{fig_GDSG_heatmap}
\vspace{-0.45cm}
\end{figure}

% \vspace{2mm}
\begin{figure}[t]
\centering
\setlength{\abovecaptionskip}{-0.1cm}
\centerline{\includegraphics[width=3.48in, height=2.76in]{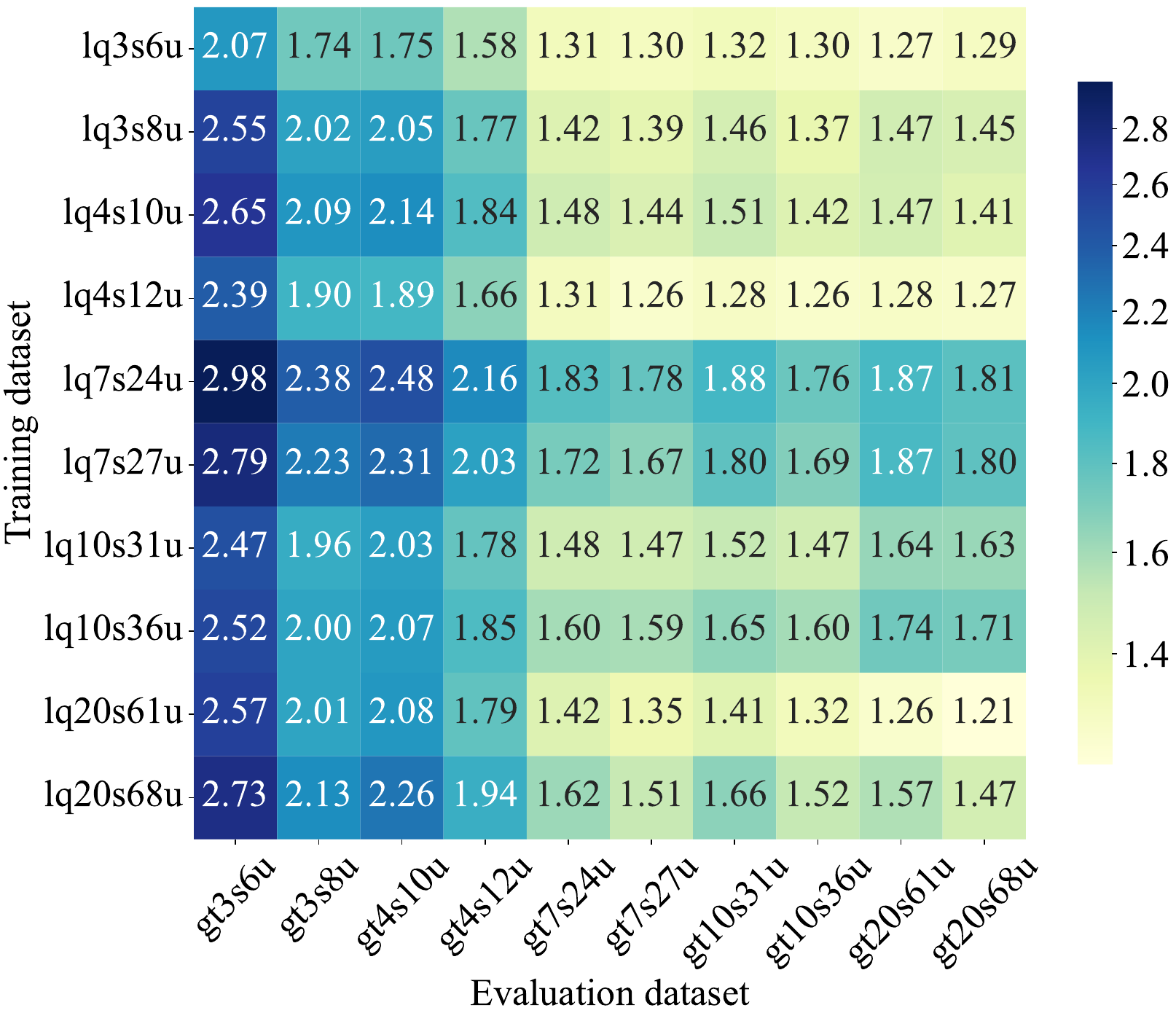}}
\caption{The performance Exceed\_ratio of DiGNN (GNN padding edge handled).}
\label{fig_DiGNN_heatmap}
\vspace{-0.45cm}
\end{figure}

\begin{figure}[h]
\centering
\setlength{\abovecaptionskip}{-0.1cm}
\centerline{\includegraphics[width=3.5in, height=2.13in]{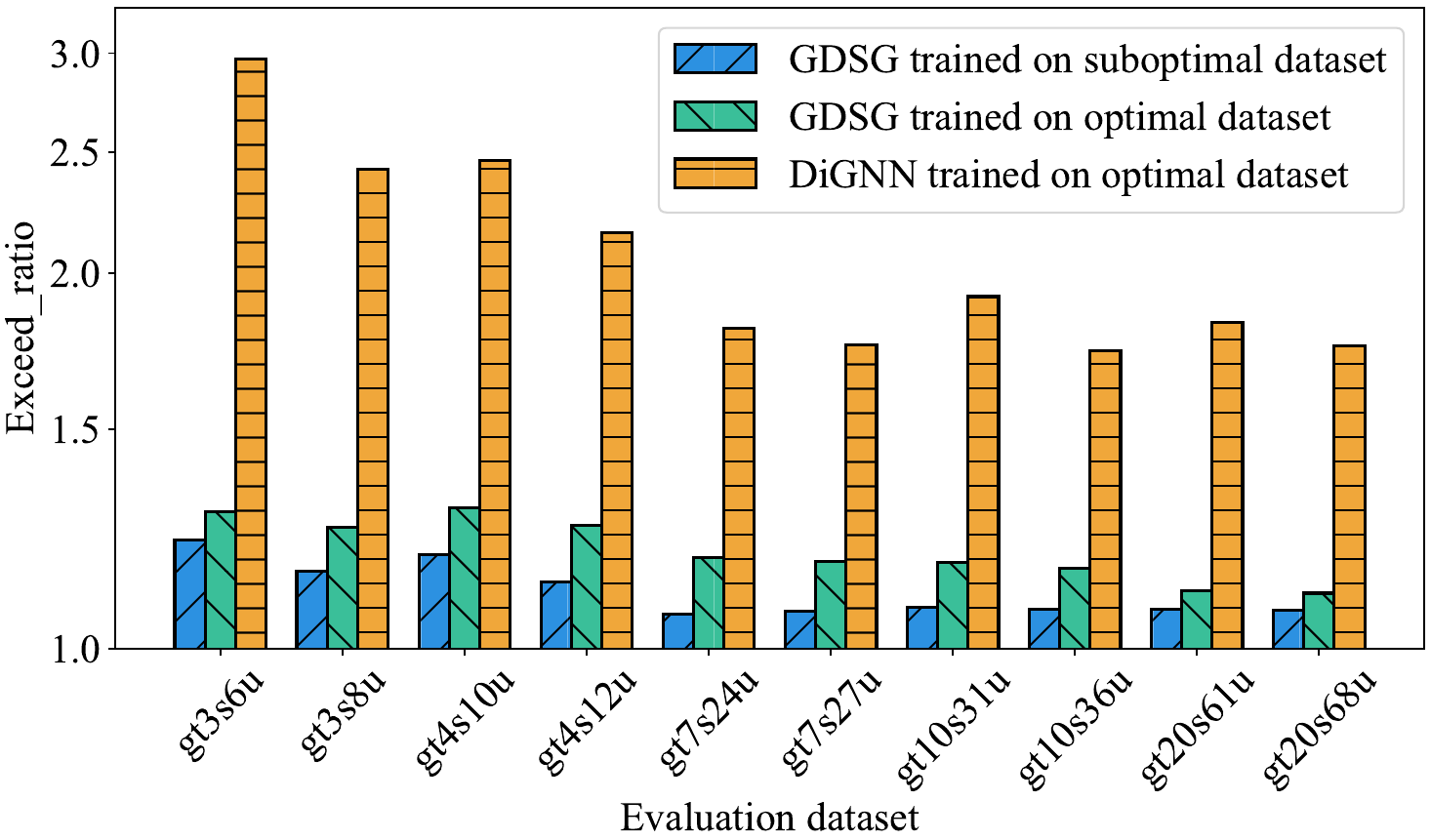}}
\caption{Performance comparison of three trained models: GDSG trained on a suboptimal dataset ($\rm lq7s24u$, 80000 samples), GDSG trained on an optimal dataset ($\rm gt3s6u$, 2000 samples) and DiGNN trained on an optimal dataset ($\rm gt3s6u$, 2000 samples).}
\label{fig_gt_train_comp}
\vspace{-0.45cm}
\end{figure}

% \vspace{1mm}
\subsection{GDSG versus DiGNN on Effectiveness and Generalization}\label{sec_vs_gnndi}
This section compares GDSG and DiGNN to illustrate the improvement brought by diffusion generation. 
We train $10$ GDSG models and $10$ DiGNN models on the $10$ suboptimal training sets and evaluate them on the $10$ optimal testing sets. In Fig. \ref{fig_GDSG_heatmap}, GDSG's $\rm Exceed\_ratio$ on most testing sets is below $1.1$, with a few reaching $1.2$, indicating near-optimal performance. This significantly outperforms DiGNN, which has a minimum $\rm Exceed\_ratio$ over $1.2$ (Fig. \ref{fig_DiGNN_heatmap}). Both models perform worse on scales with $3$ and $4$ servers due to the denser input graphs causing over-smoothing \cite{dwivedi2023benchmarking} in deep iterative GNNs. The slightly poorer performance at the corners and edges of the heatmaps suggests some generalization issues with GNN implementation. Overall, despite some engineering flaws, GDSG trained on suboptimal datasets not only approaches the optimal solution but also surpasses the discriminative model and shows good generalization. 

We also build $4$ optimal training sets of small-scale problems, each with $2000$ samples, to train $4$ GDSG models and $4$ DiGNN models. We compare the best-performing GDSG from Fig. \ref{fig_GDSG_heatmap} with the best GDSG and DiGNN trained on the optimal dataset. Fig. \ref{fig_gt_train_comp} shows that GDSG outperforms DiGNN even with the same optimal dataset, and suboptimal training with sufficient samples is consistently better. \textcolor{black}{For the models trained and evaluated on the truth dataset $\rm gt3s6u$, GDSG can save up to $56.62\%$ of the total target cost.} This demonstrates that GDSG can effectively learn from both suboptimal and optimal samples.

\textcolor{black}{The model depicted in Fig. \ref{fig_GDSG_before_heatmap} utilizes the same neural network backbone and training methodology as DIFUSCO \cite{sun2023difusco}. The main distinction of the GDSG model, illustrated in Fig. \ref{fig_GDSG_heatmap}, is the explicit modification of the gating mechanism in the graph convolutional layers.} Comparing Fig. \ref{fig_GDSG_before_heatmap} and Fig. \ref{fig_GDSG_heatmap}, our GNN uses the generalization improvements with the maximum improvement exceeding $46.88\%$ ($1.166$ vs $2.195$ evaluated on $\rm gt3s8u$). Other generalization improvement results are available in open source. In short, the GDSG model's outstanding performance shows how effectively it combines solution space parametrization with diffusion generative learning on suboptimal training sets.

\begin{table*}[t]
    \centering
    \setlength{\abovecaptionskip}{-0.05cm}
    \caption{Performance comparison of Exceed\_ratio between GDSG and baselines.}
    \begin{tblr}{
      colspec={X[0.85,l] X[0.5,c] X[0.75,c] X[0.45,c] X[0.45,c] X[0.55,c] X[0.4,c]},
      cells = {c},
      hlines, vlines,
      rowsep = {2pt},
      row{1} = {0ex},
    }
    \textbf{Evaluation Dataset} & \textbf{MTFNN} & \textbf{HEU\ \ \ \ \ \ \ \ \ \ \ \ \ \ \ \ \ \ \ \ \ \ \ \textcolor{black}{(dataset generator)}} & \textbf{DiGNN} & \textbf{DIFUSCO} & \textbf{T2T\ \ \ \ \ \ \ \ \ \ \ \ \ \ \ \ \ \ \ \ \textcolor{black}{(not applicable)}} & \textbf{GDSG (Ours)} \\
    $\rm gt3s6u$                & 83.44          & 1.22         & 2.07           & 1.22             & -            & \textbf{1.22}        \\
    $\rm gt3s8u$                & 288.93         & 1.17         & 1.74           & 1.18             & -            & \textbf{1.15}        \\
    $\rm gt4s10u$               & 3493837.38     & 1.20         & 1.75           & 1.19             & -            & \textbf{1.19}        \\
    $\rm gt4s12u$               & 225.07         & 1.17         & 1.58           & 1.15             & -            & \textbf{1.13}        \\
    $\rm gt7s24u$               & 804.12         & 1.13         & 1.31           & 1.10             & -            & \textbf{1.07}        \\
    $\rm gt7s27u$               & 407798.51      & 1.14         & 1.30           & 1.09             & -            & \textbf{1.07}        \\
    $\rm gt10s31u$              & 2672.08        & 1.14         & 1.32           & 1.10             & -            & \textbf{1.08}        \\
    $\rm gt10s36u$              & 230.47         & 1.15         & 1.30           & 1.08             & -            & \textbf{1.08}        \\
    $\rm gt20s61u$              & 1272.67        & 1.15         & 1.27           & 1.12             & -            & \textbf{1.08}        \\
    $\rm gt20s68u$              & 4517.63        & 1.18         & 1.29           & 1.10             & -            & \textbf{1.07}        
    \end{tblr}
    \label{tab_baselines}
    \vspace{-0.45cm}
\end{table*}

% \vspace{1mm}
\subsection{Comparison with Baseline Schemes}
We compare the $\rm Exceed\_ratio$ of GDSG with all baselines in Table \ref{tab_baselines}. MTFNN, which can only process vector data and lacks graph understanding capability, can train only one model per test scale, resulting in very poor performance. In some problems, MTFNN's $\rm Exceed\_ratio$ reaches tens of thousands due to the large data volume and high communication costs of computationally intensive tasks. Incorrect offloading by MTFNN leads to extremely high total cost. In comparison, HEU and DiGNN can achieve relatively better results, and GDSG also outperforms HEU and DiGNN across the board. Although HEU uses the MCMF algorithm to ensure the optimal offloading decision for a given computational resource allocation, its heuristic initialization of resource allocation cannot guarantee optimality, leading to suboptimal results. Furthermore, GDSG reduces the total target cost by up to $41.06\%$ (GDSG trained on $\rm lq7s24u$ vs DiGNN trained on $\rm lq3s6u$ evaluated on gt3s6u). 

DIFUSCO performs relatively worse than GDSG, with the performance gap primarily due to the lack of consideration for padding edges in the GNN. Moreover, T2T introduces automatic gradient computation of the target optimization function's gradient during the denoising process. While this is computable for TSP and MIS problems in the original work, our computation offloading objective function is non-differentiable, making gradients unavailable. As a result, T2T is not applicable to MEC optimization problems. 

\subsection{\textcolor{black}{Complexity Analysis}}
\textcolor{black}{To facilitate the application of the GDSG model, we present a complexity analysis. Our model primarily features a graph convolutional neural network with a gating mechanism \cite{dwivedi2023benchmarking}. It incorporates input layers for edge feature embeddings, node feature embeddings, and time embeddings, along with output layers designed for both continuous and discrete solutions.}

\textcolor{black}{Let \( V \) and \( E \) denote the number of nodes and edges in the input graph, respectively. Let \( S \) as the number of graph convolutional modules, \( h \) as the hidden vector dimension, \( d_v \) as the input node feature dimension, and \( d_e \) as the input edge feature dimension. The time complexity and space complexity of the temporal embedding input layer are $O(h+\frac{3}{4}h^2)$ and $O(2h)$, respectively. For the node feature embedding layer, the time and space complexity are $O(d_vh+h^2)$ and $O(d_v+h)$. Similarly, the time and space complexity of the edge feature embedding layer are $O(d_eh+h^2)$ and $O(d_e+h)$.} \textcolor{black}{For a single graph convolutional module, the time complexity and space complexity are $O(8h^2+2V+E)$ and $O(6h)$, respectively. During the graph convolution process, the time embedding layer has a time complexity of $O(\frac{1}{2}h^2)$ and a space complexity of $O(\frac{1}{2}h)$, while each module's output layer has a time complexity of $O(E+h^2)$ and a space complexity of $O(E+h)$.  }

\textcolor{black}{By omitting constants and simplifying based on $d_v,d_e\ll h$, the final time and space complexity of the neural network model are $O(h^2S+VS+ES)$ and $O(hS+ES)$, respectively. Considering the total solution generation process with \( T \) diffusion steps, the overall time complexity is $O((h^2S+VS+ES)T)$. Since \( T \) can be regarded as a small constant when applying DDIM \cite{song2021denoising}, this remains computationally efficient.} \textcolor{black}{Moreover, GDSG's running time scales with problem size, taking about $0.4$ seconds per inference with 16 parallel samplings in the $\rm gt20s68u$ test and $0.03$ seconds with 16 parallel samplings in the $\rm gt3s6u$ test. This is comparable to other common deep learning methods, meeting the real-time requirements for network optimization.}

\section{Conclusion}
%In this study, we introduce GDSG, a model for generating solutions based on graph diffusion, especially delivering optimization solution generation capabilities that are robust to training data. We focus on NP-hard optimization problems in MEC networks that lack efficient approximation algorithms, using the widely prevalent multi-server multi-user computation offloading problem as a concrete scenario. Our model converts network optimization problems into distribution learning through a defined parameterization of the solution space, which supports learning from suboptimal training sets that can be obtained efficiently. The paper provides important findings and explanations that advocate the application of generative models in solution generation. This application offers a cohesive problem formulation and model framework for network optimization. Furthermore, we construct a database for the multi-server multi-user computation offloading (MSCO) optimization problem, including suboptimal training sets, optimal training sets for a small number of small-scale problems, and optimal test sets. Through our experiments, GDSG has shown superior performance compared to baseline methods, whether on the suboptimal training set or the optimal training set. The proposed GDSG approach disrupts the current trend in learning-based network optimization methods that rely heavily on extensive ground-truth datasets, thereby reducing the reliance on real optimal datasets. 

In this study, we have introduced GDSG, a model designed for generating solutions based on graph diffusion, particularly focused on solving optimizing problems that remain robust to the limitations of training data. Our primary emphasis is on NP-hard optimization problems in MEC networks that currently lack efficient approximation algorithms, with the multi-server multi-user computation offloading (MSCO) problem serving as a specific example.
The proposed GDSG model transforms network optimization problems into a framework of distribution learning by systematically parameterizing the solution space. This allows us to efficiently learn from suboptimal training sets, which can be more easily gathered compared to optimal ones. Our findings highlight the potential of generative models in solution generation, offering a unified structured model framework for solving network optimization problems. \textcolor{black}{The results of GDSG highlight a pathway that avoids the high complexity constraints of existing numerical algorithms while significantly enhancing the robustness of training data and the application efficiency of AI-based methods.}

Additionally, we have developed a comprehensive database for the MSCO optimization problem. This includes suboptimal training sets, a limited number of optimal training sets for smaller-scale problems, and optimal testing sets. \textcolor{black}{With the current experimental results, we believe that cross-scale generalization and model simplification through distillation could be two promising directions for future research.}

%We identify several future directions, one of which involves developing more explicit mechanisms to ensure that the generated solutions adhere to the constraints inherent in optimization problems. Additionally, the presence of numerous unknown variables within the theoretical framework indicates that there is still room for further exploration regarding comprehensive conclusions on convergence and the requirements for data. Addressing these aspects could significantly enhance the robustness and applicability of the GDSG model.

%We have considered some future directions of this work, one of which is ensuring that generated solutions satisfy the constraints of optimization problems through more explicit mechanisms. Additionally, the presence of numerous unknown variables in the theoretical framework means that comprehensive conclusions on convergence or data requirements remain room for further investigation. 
\vspace{-2mm}

% \section*{Acknowledgments}
% This should be a simple paragraph before the References to thank those individuals and institutions who have supported your work on this article.

% {\appendix[Proof of the Zonklar Equations]
% Use $\backslash${\tt{appendix}} if you have a single appendix:
% Do not use $\backslash${\tt{section}} anymore after $\backslash${\tt{appendix}}, only $\backslash${\tt{section*}}.
% If you have multiple appendixes use $\backslash${\tt{appendices}} then use $\backslash${\tt{section}} to start each appendix.
% You must declare a $\backslash${\tt{section}} before using any $\backslash${\tt{subsection}} or using $\backslash${\tt{label}} ($\backslash${\tt{appendices}} by itself
%  starts a section numbered zero.)}

%{\appendices
%\section*{Proof of the First Zonklar Equation}
%Appendix one text goes here.
% You can choose not to have a title for an appendix if you want by leaving the argument blank
%\section*{Proof of the Second Zonklar Equation}
%Appendix two text goes here.}

\bibliographystyle{IEEEtran}  
%\bibliography{references}
% Generated by IEEEtran.bst, version: 1.14 (2015/08/26)

% \newpage

% \section{Biography Section}
% If you have an EPS/PDF photo (graphicx package needed), extra braces are
%  needed around the contents of the optional argument to biography to prevent
%  the LaTeX parser from getting confused when it sees the complicated
%  $\backslash${\tt{includegraphics}} command within an optional argument. (You can create
%  your own custom macro containing the $\backslash${\tt{includegraphics}} command to make things
%  simpler here.)
 
% \vspace{11pt}

% \bf{If you include a photo:}\vspace{-33pt}
% \begin{IEEEbiography}[{\includegraphics[width=1in,height=1.25in,clip,keepaspectratio]{fig1}}]{Michael Shell}
% Use $\backslash${\tt{begin\{IEEEbiography\}}} and then for the 1st argument use $\backslash${\tt{includegraphics}} to declare and link the author photo.
% Use the author name as the 3rd argument followed by the biography text.
% \end{IEEEbiography}

% \vspace{11pt}

% \bf{If you will not include a photo:}\vspace{-33pt}
% \begin{IEEEbiographynophoto}{John Doe}
% Use $\backslash${\tt{begin\{IEEEbiographynophoto\}}} and the author name as the argument followed by the biography text.
% \end{IEEEbiographynophoto}

\vfill

\end{document}